\documentclass[3p,12pt]{elsarticle}
\usepackage{lineno,hyperref}
\usepackage{amsmath,bm,geometry,amssymb}
\usepackage{amsfonts}
\usepackage{color}
\usepackage{nomencl,etoolbox,ragged2e,siunitx}
\usepackage{mathtools}

\usepackage{multirow}
\usepackage{caption}
\usepackage{parskip}
\usepackage{calligra}
\usepackage{makeidx}
\usepackage{tikz}
\usepackage{subcaption}
\usepackage[utf8]{inputenc}
\usepackage[english]{babel}
\usepackage{amsthm}

\usetikzlibrary{matrix}
\makeindex
\DeclareMathAlphabet{\mathcalligra}{T1}{calligra}{m}{n}
\begin{document}
	\title{General Implicit Iterative Method for Unified Gas-kinetic Scheme}
	\author[ad1]{Xiaocong Xu}
	\ead{xxuay@connect.ust.hk}
	\author[ad1]{Yajun Zhu}
	\ead{mazhuyajun@ust.hk}
	\author[ad2]{Chang Liu}
	\ead{liuchang@iapcm.ac.cn}
	\author[ad1,ad3]{Kun Xu\corref{cor1}}
	\ead{makxu@ust.hk}
	\address[ad1]{Department of Mathematics, The Hong Kong University of Science and Technology, Hong Kong, China}
	\address[ad2]{Institute of Applied Physics and Computational Mathematics, No. 2, FengHao East Road, HaiDian District, Beijing 100094, China}
	\address[ad3]{HKUST Shenzhen Research Institute, Shenzhen 518057, China}
	\cortext[cor1]{Corresponding author}
	
	\begin{abstract} 	 			
In order to further enhance the computational efficiency of the implicit unified gas-kinetic scheme (IUGKS, JCP 315 (2016) 16-38) for multi-scale flow simulation, a two-step IUGKS is proposed in this paper. The multiscale solution of the UGKS is determined by the integral solution of the kinetic model equation, which is composed of the Lagrangian integration of the equilibrium and the free particle transport of the nonequilibrium state. With the implicit evaluation of the macroscopic variables in the first iterative step, the integration of the equilibrium can be directly used in the flux calculation of macroscopic flow variables in the second iterative step. This is equivalent to include the viscous flux in the implicit scheme to accelerate the convergence of the solution instead of using the Euler flux in the iterative process of the original IUGKS. In the present IUGKS, the update of macroscopic flow variables are closely coupled with the implicit evolution of the gas distribution function.
At the same time, in order to get the more accurate solution, the full Boltzmann collision term is incorporated into the current scheme through the penalty method. Different iterative techniques, such as LU-SGS and multi-grid, are used for solving the linear algebraic system of coupled macroscopic and microscopic equations. The efficiency of the IUGKS has reached a favorable level among all implicit schemes for the kinetic equations in the literature. 
Several numerical examples are used to validate the performance of the IUGKS. Accurate solutions have been obtained efficiently in all flow regimes from low speed to hypersonic ones.
	\end{abstract}
	\begin{keyword}
		unified gas-kinetic scheme; implicit iterative method; multigrid; Boltzmann collision operator
	\end{keyword}
	\maketitle

	\section{Introduction}
	The gas dynamics can be modeled with a variable scale in different flow regimes.
	For the continuum flow, the well-known macroscopic governing equations for fluid dynamics are the Euler and Naver-Stokes (NS) equations.
	While in the rarefied regime, the kinetic equation on the particle mean free path scale becomes a reliable and effective description for non-equilibrium flow.
	The Boltzmann equation \cite{boltzmann2012lectures} is the fundamental equation for rarefied gas dynamics.
	Due to the complexity of collision operator in the Boltzmann equation, some relaxation models, such as the Bhatnagar--Gross--Krook (BGK) model \cite{BGK1954}, the ellipsoidal statistical BGK (ES-BGK) model \cite{holway1966new}, and the Shakhov BGK (S-BGK) model \cite{shakhov1968generalization}, have been developed and are commonly used in both academic research and engineering application.
	Many kinetic solvers have been constructed for flow simulation and most of them are required to have kinetic scale resolution in order to
provide valid solutions.
	While, in the continuum regime, the kinetic method usually has poor performance due to the stiffness of collision term and huge numerical discretization, and the hydrodynamic flow solvers with the evolution of macroscopic flow variables are more accurate and efficient.
	The Chapman-Enskog procedure \cite{chapman1990mathematical} gives the quasi-equilibrium distribution function in the continuum regime, which can be used directly in the construction of a hydrodynamic flow solver, such as the gas-kinetic scheme (GKS) for the Navier-Stokes solutions \cite{xu2001}.
	To simulate flows in all regimes, the unification of both kinetic and hydrodynamic modelings becomes necessary in the multiscale method \cite{xu-book}.

	To solve the kinetic equations, there are mainly two types of numerical methods, i.e., the stochastic methods and the deterministic methods.
	The direct simulation Monte Carlo (DSMC) \cite{bird1994molecular} is the most popular stochastic method for high-speed rarefied flow.
	However, the particle method suffers from statistical noise, which makes it difficult in the low-speed flow simulation.
	The discrete velocity method (DVM) \cite{mieussens2000discrete}, as a deterministic method, is absent from noise, but becomes
	extremely expensive due to the discretization of particle velocity space, especially for the high speed and high temperature flow.
	Both DSMC and conventional DVM are based on the operator splitting treatment for the particle transport and collision.
	In order to keep the accuracy, the cell-size related numerical dissipation has to be properly controlled.
	In the above DSMC and DVM,  the cell size and time step are restricted to be less than the particle mean free path and collision time in the explicit numerical evolution process. 	Under such a constraint, the computational cost will increase rapidly for the near continuum and continuum flow simulations.
	In order to overcome this difficulty, based on the micro-macro decomposition or the penalty technique, many asymptotic preserving (AP) \cite{jin1999efficient} schemes have been proposed and developed to capture the Euler/Navier-Stokes limits.
	Based on the direct modeling on the cell size and time step scale, the unified gas-kinetic scheme (UGKS) \cite{xu2010unified, huang2012,liu2014unified,liu2016,liu2017} has been proposed for multi-scale simulation in all flow regimes. The key ingredient of the UGKS is that the numerical flux across the cell interface is constructed from the integral solution of the kinetic model equation, which couples the particle transport and collision in an evolution process and uses the accumulating solution of particles' collision within a numerical time step for the update of the solution. In other words, the local flow physics recovered by the UGKS is determined by the ratio of particle collision time and the numerical time step, i.e., the cell's Knudsen number.
	The UGKS has an asymptotic limit to the NS solution in the continuum flow regime without the kinetic scale restriction on the time step and cell size.

	For the steady state calculation, many implicit schemes for the kinetic equations have been developed for fast convergence.
	The high-order and low-order (HOLO) method \cite{taitano2014moment, chacon2017multiscale} couples the high-order kinetic equation and a low-order fluid moment equation, targeting on fast convergence in both continuum and rarefied regimes.
	With a similarly coupled system, the recent-developed general synthetic iterative scheme (GSIS) \cite{su2020can, zhu2021general} shows a promising efficiency in multi-scale flow simulation, where the constitutive relations of NS shear stress and heat conduction are explicitly used
	to guide the convergence of macroscopic solutions.
	For the UGKS, standard implicit techniques in CFD, such as LU-SGS and multigrid, have been incorporated in IUGKS \cite{zhu2016implicit, zhu2017unified, zhu2019implicit}, which made a significant improvement in convergence for the steady state solution.
	The present work is targeting on the further improvement of IUGKS, where an even closer coupling of both macroscopic and microscopic evolution equations will be developed in an implicit way. 
By taking the NS dissipative terms into account, a two-step IUGKS will be constructed and be equipped with a multi-grid solver. 
A comparative study of  efficiency will be presented from different  algebraic iterative methods.
	In order to treat the collision term in a fully implicit way, the macroscopic governing equations will be solved implicitly first for the prediction of the equilibrium state. Afterwards, based on the predicted conservative flow variables, the evolution equation of the distribution function forms a diagonal matrix system. 
	Then, both the matrix systems derived from implicit macroscopic equations and implicit microscopic equations are solved iteratively using the Lower-upper symmetric Gauss-Seidel (LU-SGS) method with a multi-grid acceleration.
	In addition, in order to achieve a reliable physical solution in the highly rarefied regime, the full Boltzmann collision operator is integrated into the IUGKS for the steady state solutions. 
	
	The paper is organized as follows. In Section~\ref{UGKS}, the explicit UGKS scheme is briefly introduced.
	In Section~\ref{IUGKS}, a general framework of implicit UGKS is presented and described in detail, which includes
	the discretization of the full Boltzmann collision operator and the two-step acceleration technique.
	The numerical procedures for solving algebraic systems raised from IUGKS for the evolution of microscopic and macroscopic equations
	will be discussed in Section~\ref{Numeric}. The numerical tests, including low speed Couette flow, the Fourier flow, lid-driven cavity flow, and the high-speed flow around a square cylinder, will be presented in Section~\ref{example}.
	The conclusion will be drawn in Section~\ref{conclusion}.
	
	\section{Kinetic equation and UGKS}\label{UGKS}
	
	\subsection{Kinetic equation}
	
	The kinetic equations describe the time evolution of the probability density function (PDF) or velocity distribution function on the kinetic scale.
	The kinetic equation models both particle transport and  collision in the following form
	\begin{equation}
		\partial _t f + \vec{u} \cdot \nabla_{\vec{x}} f = Q,
		\label{KinEq}\end{equation}
	where $f = f(\vec{x}, t, \vec{u})$ is the velocity PDF, $\vec{x}$ is the spatial position, $t$ is the time, and $\vec{u}$ is the particle velocity.
The collision term $Q$ can be modeled by the nonlinear Boltzmann collision term $Q(f,f)$ and the kinetic relaxation process.
The Boltzmann collision term takes the form
	\begin{equation}\label{Boltzmanncollision}
		Q(f,f)(\vec{u}) = \int_{\mathbb{R}^3}\int_{\mathbb{S}^{2}}B(|\vec{u}-\vec{u}_{*}|,\sigma) \left[ f(\vec{u}'_{*})f(\vec{u}') - f(\vec{u}_{*})f(\vec{u})\right]d\sigma d\vec{u}_{*}.
	\end{equation}
where $B(|\vec{u}-\vec{u}_{*}|,\sigma)$ is the collision kernel.
The numerical evaluation of the nonlinear Boltzmann collision operator is highly expensive due to its five-fold integration form. In order to simplify the formulation, the relaxation type of collision operator is considered here, such as the S-BGK model
	\begin{equation}
		Q(f) = \frac{g - f}{\tau},
	\end{equation}
	where $\tau$ is the relaxation parameter. The equilibrium state $g$ takes the form of
	\begin{equation}
		g = g^M\left[ 1 + (1 - \mathrm{Pr})\vec{c}\cdot \vec{q}\left(\frac{\vec{c}^2}{RT} - 5 \right)\right],
	\end{equation}
and $g^M$ is the Maxwellian distribution
\begin{equation}
g^M = \frac{\rho}{(2\pi RT)^{3/2}}\mathrm{exp}\left( -\frac{\vec{c}^2}{2RT}\right)	
\end{equation}
	where $\rho$ is the mass density, $R$ is the specific gas constant, $\mathrm{Pr}$ is the Prandtl number, $\vec{c} = \vec{u} - \vec{U}$ is the peculiar velocity with $\vec{U}$ the macroscopic flow velocity, and $\vec{q}$ is the heat flux.
	The relation between probability density function and macroscopic conservative flow variables are defined as
	\begin{equation}
		\vec{W} = \int f\vec{\psi} d\vec{u}.
		\label{moments}\end{equation}
	Here, $\vec{\psi} = \left[ 1, \vec{u}, \frac{1}{2}|\vec{u}|^2\right]^{T}$ is the vector of collision invariants and $\vec{W} = \left[\rho, \rho \vec{U}, \rho E \right]^T$ is the vector of conservative flow variables. The pressure tensor $\mathbf{P}$ and the heat flux $\vec{q}$ can be also calculated from the PDF by
	\begin{equation}
			\mathbf{P} = \int \vec{c} \vec{c} f d\vec{u}, \quad
			\vec{q} = \frac{1}{2} \int \vec{c} |\vec{c}|^2f d\vec{u}.
	\end{equation}
	Based on the kinetic equation, a unified gas kinetic scheme has been developed \cite{xu2010unified}, which accurately captures both the kinetic and hydrodynamic solution in the corresponding regime.
	\subsection{Unified gas-kinetic scheme}
	In this subsection, the explicit UGKS will be introduced.
	The UGKS takes direct modeling of flow physics on the discretization scales, i.e., the time step and cell size.
	In the framework of a finite volume method, if the trapezoidal rule is used for the collision term, the conservation of the gas distribution function can be described as
	\begin{equation}
		f_{i,k}^{n+1} = f_{i,k}^n - \frac{1}{V_i} \sum _{j \in N(i)} S_{ij} \int_{t^n}^{t^{n+1}} u_{k,n}f_{ij,k}(t) dt + \frac{\Delta t}{2} \left( \frac{g_{i,k}^n - f_{i,k}^n}{\tau^n_i} + \frac{g_{i,k}^{n+1} - f_{i,k}^{n+1}}{\tau^{n+1}_i} \right),
		\label{Discre}\end{equation}
	where $f_{i,k}^{n}$ and $f_{i,k}^{n+1}$ are spatially averaged probability density functions of cell $i$ at particle velocity $\vec{u}_k$ and at time $t^n$ and $t^{n+1}$, respectively. $N(i)$ is the index set of the neighbors of cell $i$, and $ij$ denotes the interface between cells $i$ and $j$. $V_i$ is the volume of cell $i$, $S_{ij}$ is the area of the interface $ij$, $u_{k,n}$ is the normal projection of paricles velocity $\vec{u}_k$ along the cell interface, and $f_{ij,k}(t)$ is a time-dependent distribution function on the interface $ij$ at particle velocity $\vec{u}_k$.
	
	Multiplying the collision invariants and integrating over the velocity space, the conservation laws of macroscopic flow variables can be obtained from Eq.~\eqref{Discre}
	\begin{equation}
		\vec{W}_i^{n+1} = \vec{W}_i^{n} - \frac{1}{V_i}\sum_{j\in N(i)} S_{ij} \left(\sum_k \int _{t^{n}}^{t^{n+1}} u_{k,n}f_{ij,k}(t)\vec{\psi}_k dt \right).
		\label{macroupdate}\end{equation}
	Here, $\vec{\psi}_k = \left[ 1, \vec{u}_k, \frac{1}{2} |\vec{u}_k|^2 \right]^T$. In order to evolve the above discretized equations, a time-dependent probability density function $f_{ij,k}(t)$ has to be constructed. Based on the integral solution of Eq.\eqref{KinEq}, a multi-scale $f_{ij,k}(t)$ can be modeled as,
	\begin{equation}
		\begin{aligned} f_{i j, k}(t)=f\left(\vec{x}_{i j}, t, \vec{u}_{k}\right) =\frac{1}{\tau} \int_{t^{n}}^{t} g\left(\vec{x}^{\prime}, t^{\prime}, \vec{u}_{k}\right) e^{-\left(t-t^{\prime}\right) / \tau} d t^{\prime}  +e^{-\left(t-t^{n}\right) / \tau} f\left(\vec{x}_{i j}-\vec{u}_{k}\left(t-t^{n}\right), t^{n}, \vec{u}_{k} \right), \end{aligned}
		\label{integralSol}\end{equation}
	where $\vec{x^{\prime}} = \vec{x}_{ij} - \vec{u}_k (t - t^{\prime})$ is the particle trajectory, and $\tau = \mu / p$ is the local mean collision time computed from the dynamic viscosity coefficient $\mu$ and the pressure $p$.
	
	For a second-order spatial accuracy, the initial probability density function around the cell interface $ij$ can be approximated by
	\begin{equation}
		f_{0,k}\left(\vec{x}\right) = f\left( \vec{x}, t^n, \vec{u}_k\right) =\left\{\begin{array}{ll}f_{i j, k}^{i}+\sigma _{i}^{n} \cdot \vec{x}, & \vec{n}_{i j} \cdot \vec{x}<0 \\ f_{i j, k}^{j}+\sigma _{j}^{n} \cdot \vec{x}, & \vec{n}_{i j} \cdot \vec{x} \geq 0\end{array}\right. ,
		\label{constructf}\end{equation}
	where $f_{ij,k}^{i}$ and $f_{ij,k}^{j}$ are the reconstructed initial probability density functions on both sides of the cell interface $ij$, $\vec{n}_{ij}$ is the unit normal vector of the cell interface $ij$, and $\sigma =\nabla_{\vec{x}}{f}$ is the spatial gradient of the initial probability density function. In addition, the equilibrium state is approximated by Taylor expansion,
	\begin{equation}
		g_{k}\left(\vec{x},t\right) = g\left(\vec{x},t, \vec{u}_k \right) = g_{0,k} +  \vec{x} \cdot \nabla_{\vec{x}}g_{0,k}  + t \partial_t g_{0,k},
		\label{constructg}\end{equation}
	where $g_{0,k}$ is the initial equilibrium at the cell interface $ij$, and $ \nabla_{\vec{x}}g_{0,k}$ and $\partial_t g_{0,k}$ are the spatial and temporal gradients of the initial equilibrium state. The detailed calculation for these terms can be found in previous literature \cite{xu2010unified}.
	Applying Eqs.~\eqref{constructf} and \eqref{constructg} to Eq.~\eqref{integralSol}, the time-dependent probability density function $f_{ij,k}(t)$ can be computed by
	\begin{equation}
		f_{ij,k}(t) = q_1 g_{0,k} + q_2 \vec{u}_k \cdot \nabla_{\vec{x}}g_{0,k} + q_3 \partial _t g_{0,k} + q_4 f_{0,k} + q_5  \vec{u}_k \cdot \nabla_{\vec{x}}f_{0,k},
		\label{fij}\end{equation}
	where the coefficients $q$ are
	\begin{equation}
		\begin{aligned}
			q_1 &= 1 - e^{-t / \tau}, \\
			q_2 &= \tau \left( e^{-t/\tau} - 1\right) + te^{-t/\tau} ,\\
			q_3 &= t - \tau + \tau e^{-t /\tau}  ,\\
			q_4 &= e^{-t / \tau},\\
			q_5 &= - te^{-t/ \tau}.
		\end{aligned}
	\end{equation}
	Rewrite Eq.~\eqref{fij} into the following form
	\begin{equation}
		\begin{aligned}
			f_{ij,k}(t) &= g_{ij,k} + q_2 \vec{u}_k \cdot \nabla_{\vec{x}}g_{ij,k} + q_3 \partial _t g_{ij,k} + q_4 \left( f_{ij,k} - g_{ij,k}\right) + q_5  \vec{u}_k \cdot \nabla_{\vec{x}}f_{ij,k},  \\
			& \triangleq \tilde{g}_{ij,k} + \tilde{f}_{ij,k},
		\end{aligned}
		\label{interfaceF}\end{equation}
	where $\tilde{g}_{ij,k}$ is the first three terms related to the equilibrium state, which may give the NS distribution function from the integration of the equilibrium state.  $\tilde{f}_{ij,k}$ is the last two terms related to the non-equilibrium state. Note that under this arrangement, the flux related to $\tilde{f}_{ij,k}$ will mainly contribute to the non-equilibrium transport, which plays an important role in
the rarefied regime. The above formulation will be used in the construction of the implicit UGKS.
	
	As long as the time-dependent probability density function $f_{ij,k}(t)$ is obtained, the flux term in Eq.~\eqref{Discre} and Eq.~\eqref{macroupdate} can be computed. Then we can update the averaged probability density function $f_{i,k}^{n+1}$ and macroscopic flow variables $\vec{W}_i^{n+1}$ in each cell.
	
	\section{General framework of the implicit UGKS}\label{IUGKS}
For steady state calculation, a finite volume implicit dicretization of Eq.~\eqref{KinEq} can be written as
	\begin{equation}
		\frac{f_{i,k}^{n+1} - f_{i,k}^{n}}{\Delta t} + \frac{1}{V_i}\sum_{j\in N(i)}S_{ij}u_{k,n}f_{ij,k}^{n+1} = \frac{\tilde{g}_{i,k}^{n+1} - f_{i,k}^{n+1}}{\tau_i^{n+1}}.
		\label{implicitDis}\end{equation}
	Here, the implicit numerical time step $\Delta t$ is introduced to improve the stability of the scheme.
	In the above equation, a prediction step should be carried out first to obtain an approximate equilibrium state $\tilde{g}^{n+1}$ so that the collision term can be treated implicitly.
	For simplicity, the formulations in this section are presented for two-dimensional cases, and the extension to three-dimensional cases would be straightforward.
	
	\subsection{Prediction step for $\tilde{g}^{n+1}$}
	Taking moments of Eq.~\eqref{implicitDis}, we can obtain the implicit governing equations of macroscopic flow variables, i.e.,
	\begin{equation}
		\frac{\tilde{\vec{W}}_{i}^{n+1} - \vec{W}_i^n}{\Delta t} + \frac{1}{V_i}\sum_{j\in N(i)}S_{ij}\vec{F}_{ij}^{n+1} = \vec{0},
	\end{equation}
	where $\vec{F}_{ij}^{n+1}$ are the fluxes for conservative variables at  the interface $ij$. Subtracting the fluxes at time $t^{n}$ from both sides, the governing equations could be rewritten as
	\begin{equation}
		\frac{1}{\Delta t}\Delta \vec{W}_{i}^{n+1}  + \frac{1}{V_i}\sum_{j\in N(i)}S_{ij}\Delta\vec{F}_{ij}^{n+1} = -\frac{1}{V_i}\sum_{j\in N(i)}S_{ij}\vec{F}_{ij}^{n},
		\label{deltaform}\end{equation}
	where $\Delta \vec{W}_{i}^{n+1} = \tilde{\vec{W}}_{i}^{n+1} - \vec{W}_i^n$ and $\Delta \vec{F}_{ij}^{n+1} = \vec{F}_{ij}^{n+1} - \vec{F}_{ij}^{n}$.
	In order to solve Eq.~\eqref{deltaform}, the fluxes on the left hand side will be approximated by the Euler equation-based fluxes, which will lead to a matrix-free algorithm,
	\begin{equation}
		\Delta \vec{F}_{ij}^{n+1} = \frac{1}{2} \left[ \vec{T}_i^{n+1} +  \vec{T}_j^{n+1} + \Gamma_{ij} \left( \vec{W}_i^{n+1} - \vec{W}_j^{n+1}\right)\right] - \frac{1}{2} \left[ \vec{T}_i^{n} +  \vec{T}_j^{n} + \Gamma_{ij} \left( \vec{W}_i^{n} - \vec{W}_j^{n}\right)\right],
	\end{equation}
	where $\vec{T}$ is the Euler flux, which can be determined by the conservative variables
	\begin{equation}
		\vec{T} =\left[\begin{array}{c}\rho U_x  \\ \rho U_{x}^2+ p \\ \rho U_{x} U_{y}+p\\ (\rho E + p) U_x \end{array}\right] n_x + \left[\begin{array}{c}\rho U_y  \\ \rho U_{x}U_{y}+ p \\ \rho U_{y}^2+p\\ (\rho E + p) U_y \end{array}\right]n_y.
	\end{equation}
	Here, $\left[n_x,n_y\right]^{T}$ is the unit normal vector along the interface, $p$ is the pressure. The factor $\Gamma_{ij}$ satisfies
	\begin{equation}
		\Gamma_{ij} \geq \Lambda_{i j} =\left|\vec{U}_{i j} \cdot \vec{n}_{i j}\right|+a_{s},
	\end{equation}
	where $\Lambda_{ij}$ represents the spectral radius of the Jacobian of the Euler flux, which can be evaluated by the macroscopic velocity $\vec{U}_{ij}$ and the speed of sound $a_s$ at the interface $ij$. Moreover, as stated in the previous literatures, a stable factor $s_{ij}$ related to the viscosity coefficient can be introduced into the calculation of $\Gamma_{ij}$,
	\begin{equation}
		\Gamma_{ij} = \Lambda_{ij} + s_{ij} = \Lambda_{ij} + \frac{2 \nu }{\Delta l}.
	\end{equation}
	The above flux is the so-called Lax-Friedrichs flux, and in order to simplify the notations, we define $$\vec{A}_i\left(\vec{W}_i\right) \triangleq  \frac{\vec{W}_i}{\Delta t} +\frac{1}{V_i}\sum_{j\in N(i)}S_{ij} \vec{F}_{LF}\left(\vec{W}_i,\vec{W}_j\right) = \frac{\vec{W}_i}{\Delta t} +\frac{1}{V_i}\sum_{j\in N(i)}S_{ij}  \frac{1}{2} \left[ \vec{T}_i +  \vec{T}_j + \Gamma_{ij} \left( \vec{W}_i- \vec{W}_j\right)\right].$$
	Then the evolution equations can be rewritten as
	\begin{equation}
		\vec{A}_i\left( \tilde{\vec{W}}_{i}^{n+1}\right) =\vec{R}_i^n +  \vec{A}_i\left( \vec{W}_{i}^{n}\right) , \quad \vec{R}_i^n = -\frac{1}{V_i}\sum_{j\in N(i)}S_{ij}\vec{F}^n_{ij},
		\label{nonlinear}\end{equation}
	where $\vec{F}^n_{ij}$ are the fluxes calculated from the time-dependent probability density function $f_{ij,k}(t)$ in Eq.~\eqref{implicitDis}, which reads as
	\begin{equation}
		\vec{F}^n_{ij} = \frac{1}{\Delta t_s}\sum_k \int_0^{\Delta t_s} u_{k,n}f_{ij,k}(t)\vec{\psi}_{k}dt.
	\end{equation}
	Here, the sum over $k$ is the numerical procedure for integration over the velocity space.
	$\Delta t_s$ is a small time step to determine the local physics scale, which is calculated the same as that in the explicit UGKS by the stability condition
	\begin{equation}
		\Delta t_s = \alpha \min \frac{\Delta x}{\max |\vec{u}_k|},
	\end{equation}
	where $\Delta x$ is the mesh size and $\alpha$ is the CFL number.
	It should be noticed that the time step $\Delta t_s$ is different from the implicit numerical time step $\Delta t$.
	In the IUGKS, $\Delta t_s$ can be regarded as a cell size related time step to decide the local physical solution in the corresponding scale by adjusting the combination of equilibrium and non-equilibrium distribution function, and it is the time step to average the numerical fluxes for controlling convergent steady state solution; while $\Delta t$ is the pseudo time step to increase the convergence stability for steady state solution.
	
	At the steady state, the Lax-Friedrichs flux $\vec{A}_i\left( \cdot \right)$ on the left and right hand sdies of Eq.~\eqref{nonlinear} will be canceled, and the convergent solution will be completely controlled by
	\begin{equation}
		\frac{1}{V_i}\sum_{j\in N(i)}S_{ij}\vec{F}_{ij}^{n} = 0,
	\end{equation}
	which ensures identical steady solutions as the explicit UGKS.
	
	By solving Eq.~\eqref{nonlinear}, we can get the prediction macroscopic conservative variables $\tilde{\vec{W}}^{n+1}$, and therefore $\tilde{g}^{n+1}$ can be obtained simultaneously. The detail for solving Eq.~\eqref{nonlinear} by using a matrix-free algorithm will be given later.
	
	\subsection{Solving the microscopic system}
	With the predicted equilibrium state $\tilde{g}^{n+1}$, Eq.~\eqref{implicitDis} can be rewritten in a delta-form
	\begin{equation}
		\frac{1}{\Delta t} \Delta f^{n+1}_{i,k} + \frac{1}{V_i}\sum_{j\in N(i)} u_{k,n}S_{ij}\Delta f_{ij,k}^{n+1} = \frac{\tilde{g}_{i}^{n+1} - f_{i}^{n}}{\tilde{\tau}_{i}^{n+1}} - \frac{1}{V_i}\sum_{j\in N(i)} u_{k,n}S_{ij} f_{ij,k}^{n}.
		\label{deltaMicro}\end{equation}
	Here, $\Delta f^{n+1} = f^{n+1} - f^{n}$. $\tilde{\tau}^{n+1}$ can be also obtained from the predicted macroscopic variables. Since different algorithms to compute the implicit fluxes at the left hand side of Eq.~\eqref{deltaMicro} will not affect the final convergent solutions, the first order upwind scheme is adopted here to compute $\Delta f_{ij,k}^{n+1}$, which reads as
	\begin{equation}
		\begin{aligned} \Delta f_{i j, k}^{n+1} &=\frac{1}{2}\left(\Delta f_{i, k}^{n+1}+\Delta f_{j, k}^{n+1}\right)+\frac{1}{2} \operatorname{sign}\left(\vec{u}_{k} \cdot \vec{n}_{i j}\right)\left(\Delta f_{i, k}^{n+1}-\Delta f_{j, k}^{n+1}\right), \\ &=\frac{1}{2}\left[1+\operatorname{sign}\left(u_{k, n}\right)\right] \Delta f_{i, k}^{n+1}+\frac{1}{2}\left[1-\operatorname{sign}\left(u_{k, n}\right)\right] \Delta f_{j, k}^{n+1}. \end{aligned}
		\label{upwind}\end{equation}
	Inserting Eq.~\eqref{upwind} into Eq.~\eqref{deltaMicro}, we can get a linear system,
	\begin{equation}
		D_{i,k} \Delta f_{i,k}^{n+1} + \sum_{j \in {N(i)}}D_{j,k}\Delta f_{j,k}^{n+1} = (\text{RHS})_{i,k}^n,
		\label{linear}\end{equation}
	where
	\begin{equation}
		\begin{aligned}
			D_{i, k} &=\frac{1}{\Delta t} + \frac{1}{\tilde{\tau}_{i}^{n+1}}+\frac{1}{2V_i} \sum_{j \in N(i)} u_{k, n} S_{i j}\left[1+\operatorname{sign}\left(u_{k, n}\right)\right], \\ D_{j, k} &=\frac{1}{2V_i} u_{k, n} S_{i j}\left[1-\operatorname{sign}\left(u_{k, n}\right)\right], \\ (\text{RHS})_{i,k}^n &= \frac{\tilde{g}_{i, k}^{n+1}-f_{i, k}^{n}}{\tilde{\tau}_{i}^{n+1}}-\frac{1}{V_i}\sum_{j \in N(i)} S_{i j} u_{k, n} f_{i j, k}^{n}.
		\end{aligned}
	\end{equation}
	These terms $D_{i,k}$ and $D_{j,k}$ will be the entries of a block diagonal matrix. Once the steady state is reached, $\Delta f_{i,k}^{n+1}$ will go to zero, and the steady state should satisfy $(\text{RHS})_{i,k}^n = 0$.

	\subsection{Extension to full Boltzmann collision operator}
	Consider the nonlinear Boltzmann in the dimensionless form, which reads as
	\begin{equation}
		\frac{\partial f}{\partial t} + \vec{u} \cdot \nabla_{\vec{x}} f = \frac{1}{\mathrm{Kn}}Q(f,f).
	\end{equation}
	where the collision operator $Q$ is a bilinear operator, which can be written as,
	\begin{equation}
		Q(g,f)(\vec{u}) = \int_{\mathbb{R}^3}\int_{\mathbb{S}^{2}}B(|\vec{u}-\vec{u}_{*}|,\sigma) \left[ g(\vec{u}'_{*})f(\vec{u}') - g(\vec{u}_{*})f(\vec{u})\right]d\sigma d\vec{u}_{*}.
	\end{equation}
	Here $(\vec{u},\vec{u}_{*})$ and $(\vec{u}',\vec{u}'_*)$ are the velocity pairs before and after a collision. By the momentum and energy conservation, one can show that $(\vec{u}',\vec{u}'_*)$ can be represented in terms of $(\vec{u},\vec{u}_*)$ as
	\begin{equation}
		\vec{u}' = \frac{\vec{u} + \vec{u}_*}{2} + \frac{|\vec{u} - \vec{u}_*|}{2}\sigma, \quad  \vec{u}'_* = \frac{\vec{u} + \vec{u}_*}{2} - \frac{|\vec{u} - \vec{u}_*|}{2}\sigma ,
	\end{equation}
	with $\sigma = (\vec{u}' - \vec{u}'_*)/|\vec{u}' - \vec{u}'_*|$ varying in the unit sphere $\mathbb{S}^{2}$. The collision kernel $B(|\vec{u}-\vec{u}_{*}|,\sigma)$ is a non-negative function depending only on $|v-v_{*}|$ and the cosine of the deflection angle $\theta$. In the following numerical examples, we consider a simplified collision kernel,
	\begin{equation}
		B(|\vec{u}-\vec{u}_{*}|,\sigma) = C_{\alpha}|\vec{u} - \vec{u}_*|^{\alpha},
	\end{equation}
	which is the so-called VHS model. The penalization technique \cite{filbet2010class} is used in the construction of our numerical scheme, which can be formulated as,
	\begin{equation}\label{numerical}
\begin{aligned}
		\frac{f_{i,k}^{n+1}-f_{i,k}^{n}}{\Delta t}+ + \frac{1}{V_i}\sum_{j\in N(i)}S_{ij}u_{k,n}f_{ij,k}^{n+1}=S_{i,k}^n+ \frac{\tilde{g}_{i,k}^{n+1}-f_{i,k}^{n+1}}{\tau_i^{n+1}} , \quad S_{i,k}^n = \frac{\left[ \mathcal{Q}\left(f_i^{n},f_i^n\right)\right]_k}{\mathrm{Kn}}-\frac{g_{i,k}^{n}-f_{i,k}^{n}}{\tau_i^n}.
\end{aligned}
	\end{equation}
	This is an implicit scheme, and the collision operator is solved by the fast spectral method explicitly. In the continuum regime, the predicted $\tilde{g}_i^{n+1}$ in the penalized term could boost the convergence of the implicit scheme. In the framework of the implicit UGKS, the only difference with the above procedure is that the $(\text{RHS})_{i,k}^n$ becomes
	\begin{equation}
		(\text{RHS})_{i,k}^n = \frac{\tilde{g}_{i, k}^{n+1}-f_{i, k}^{n}}{\tilde{\tau}_{i}^{n+1}}-\frac{1}{V_i}\sum_{j \in N(i)} S_{i j} u_{k, n} f_{i j, k}^{n} + S^n_{i,k}.
	\end{equation}

	\subsection{Two-step iteration for acceleration}
	The development of new acceleration method is based on two observations. Firstly, in the implicit UGKS, the microscopic system is mainly driven by the predicted $\tilde{g}^{n+1}$. Therefore, if we can have a good prediction (closer to the steady solution), the computational cost can be much reduced, especially in the continuum regime.
	The second observation is that in the continuum regime, the UGKS flux is dominated by $\tilde{g}_{ij,k}$ in Eq.~\eqref{interfaceF}. Instead of updating the $f_{ij,k}$ in each discrete velocity point and then using numerical quadrature to get the macroscopic flux, the macroscopic flux related to the analytical part $\tilde{g}_{ij,k}$ can be computed with relatively low computational cost by using the analytical formulations.
	
	In the previous study, in the continuum regime, the left hand side of the implicit UGKS is driven by the Euler flux. Based on the multi-scale property of the UGKS flux, we can naturally take the viscosity flux into account and use the viscous solution to boost the convergence of implicit UGKS. Similar consideration can be found in Yuan \textit{et al.} \cite{yuan2021multi}.
	
	Consequently, we can take one more prediction step for $\tilde{g}^{n+1}$ in the computational procedure. More specifically,
	by the first step iteration in solving Eq.~\eqref{nonlinear} for macroscopic variables, an equilibrium $\tilde{g}^{*}$ term in Eq.~\eqref{implicitDis} can be predicted. Then, we can update the macroscopic flux by
	\begin{equation}
		\vec{F}_{ij}^{n,*} =  \frac{1}{\Delta t_s}\int_0^{\Delta t_s} \langle \vec{u}\cdot \vec{n} \vec{\psi} \tilde{g}_{ij,k}^{*} \rangle dt  + \frac{1}{\Delta t_s}\sum_k \int_0^{\Delta t_s}u_{k,n}\tilde{f}_{ij,k}^{n}\vec{\psi}_{k}dt .
	\end{equation}
	The evolution equations for the second step iteration can be written as
	\begin{equation}
		\vec{A}_i\left( \tilde{\vec{W}}_{i}^{n+1}\right) =\vec{R}_i^{n,*} +  \vec{A}_i\left( \vec{W}_{i}^{*}\right) , \quad \vec{R}_i^{n,*} = -\frac{1}{V_i}\sum_{j\in N(i)}S_{ij}\vec{F}_{ij}^{n,*} ,
		\label{Newnonlinear}\end{equation}
	where $ \vec{W}_{i}^{*}$ is the conservative variables calculated by  $\tilde{g}^{*}$.
	
	So far, a brief summary of the present implicit scheme can be given, and the major steps are listed as follows
	\begin{itemize}
		\item[1.]  Get the predicted $\tilde{g}^{n+1}$ by iterating Eq.~\eqref{nonlinear} and Eq.~\eqref{Newnonlinear} in order from initial $\vec{W}^{n}$.
		\item[2.]  Get the $\Delta f^{n+1}$ by solving Eq.~\eqref{linear} with the predicted $\tilde{g}^{n+1}$ and initial $f^{n}$.
		\item[3.]  Update the distribution function $f^{n+1}$ and get the updated macro conservative variables from Eq.~\eqref{moments}.
		\item[4.] Repeat the steps 1 to 3 until satisfying the convergence criterion.
	\end{itemize}

	\section{Iterative methods for solving coupled macro-micro system}\label{Numeric}
	An improved implicit UGKS will be presented to solve the implicit system.
	For each iteration of Eq.~\eqref{Newnonlinear}, we are actually solving Eq.~\eqref{nonlinear} with the refreshed $\vec{F}_{ij}$ and $\vec{W}_i$ at the right hand side.
	In the following, the detailed methods for solving Eq.~\eqref{nonlinear} and Eq.~\eqref{linear} will be discussed.
	\subsection{Lower-Upper Symmetric Gauss-Seidel method}
	Subtracting $\vec{A}_i(\vec{W}_i^n)$ on both sides of Eq.~\eqref{nonlinear} and perform the Taylor approximation, we can obtain
	\begin{equation}
		\frac{\partial \vec{A}}{\partial \vec{W}}\Delta \vec{W}^n = \vec{R}^n.
		\label{OriginAE}\end{equation}
	The Jacobian matrix splitting method in \cite{yoon1988lower} can be used to solve the above system. 
However, the computation of the Jacobian matrix requires additional storage. The inverse linearization can be used to eliminate the computation of the Jacobian matrix \cite{sharov1997reordering}, in which the following approximation is adopted
	\begin{equation}
		\begin{aligned}
			\frac{\partial \vec{A}_i}{\partial \vec{W}_i}\Delta \vec{W}_i^n &\approx \vec{A}_i\left(\vec{W}_i^{n} +\Delta \vec{W}_i^n \right) - \vec{A}_i\left(\vec{W}_i^{n} \right) \\
			& = \left(\frac{1}{\Delta t} + \frac{1}{2}\sum_{j \in N(i)}S_{ij}\Gamma_{ij} \right)\Delta \vec{W}^{n} + \frac{1}{2}\sum_{j \in N(i)} S_{ij}\left[ \vec{T}\left(\vec{W}_{j}^{n} + \Delta\vec{W}_j^{n} \right) - \vec{T}\left( \vec{W}_j^n\right)  -\Gamma_{ij}\Delta\vec{W}_j^n \right].
		\end{aligned}
		\label{ApproAE}\end{equation}
	Therefore, the LU-SGS method for the algebraic system Eq.~\eqref{OriginAE} and Eq.~\eqref{ApproAE} can be described by a forward step
	\begin{equation}
		\Delta \vec{W}_{i}^{*} = D_i^{-1} \left( \vec{R}_i^n - \frac{1}{2}\sum_{j \in L(i)} S_{ij}\left[ \vec{T}\left(\vec{W}_{j}^{n} + \Delta\vec{W}_j^{*} \right) - \vec{T}\left( \vec{W}_j^n\right)  -\Gamma_{ij}\Delta\vec{W}_j^* \right] \right),
	\end{equation}
	and a backward step
	\begin{equation}
		\Delta \vec{W}_i^{n} = \Delta\vec{W}_{i}^{*} - D_i^{-1} \left( \frac{1}{2}\sum_{j \in U(i)} S_{ij}\left[ \vec{T}\left(\vec{W}_{j}^{n} + \Delta\vec{W}_j^{n} \right) - \vec{T}\left( \vec{W}_j^n\right)  -\Gamma_{ij}\Delta\vec{W}_j^n \right] \right),
	\end{equation}
	where $D_i$ is the diagonal element of the matrix
	\begin{equation}
		D_i = \frac{1}{\Delta t} + \frac{1}{2}\sum_{j \in N(i)}S_{ij}\Gamma_{ij} .
	\end{equation}
	Here, $L(i)$ is the index set of the neighboring cells of cell $i$ occupying the lower triangular part of the matrix, and $U(i)$ is the index set of the neighboring cells of cell $i$ occupying the upper triangular part of the matrix. The numbering of the index is essential to the LU-SGS sweeping, and bad numbering may lead to local degeneration from Gauss-Seidel iteration to Jacobian iteration. 
	\subsection{Multi-grid method}
	In this subsection, the geometry type of multi-grid method \cite{briggs2000multigrid} for solving the macro-system \eqref{nonlinear} and micro-system \eqref{linear} will be implemented and used in the calculation of 2D numerical examples.
	
\begin{figure}
		\centering
		\includegraphics[width=0.48\textwidth]{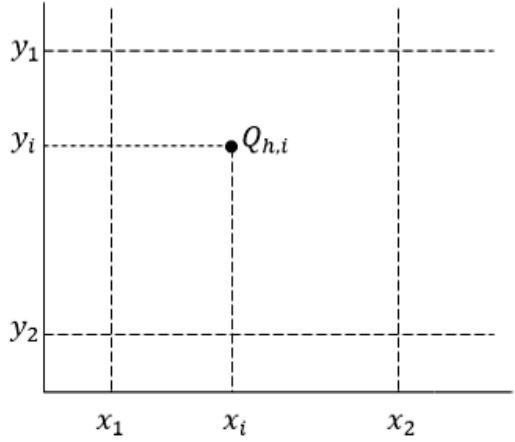}
		\caption{Bilinear interpolation}
		\label{BiIn}
	\end{figure}

	Firstly, the restriction and interpolation operator should be properly defined to connect different levels of mesh \cite{zhu2017unified}. The restriction operator $I_{h}^{H}$, which maps the quantities from a fine mesh to a coarse mesh, can be naturally defined as the average quantities of all the related cells in the fine mesh. Specifically, for a quantity $Q_h$ in a fine mesh, the restricted quantity $Q_H$ in a coarse mesh can be calculated as
	\begin{equation}
		Q_{H,i} = I_{h}^{H}(\{ Q_{h,j}, j \in S(i)\}) = \frac{\sum_{j \in S(i)}Q_{h,j}V_{h,j}}{\sum_{j \in S(i)}V_{h,j}},
	\end{equation}
	where the subscripts $i,j$ denote the quantities in cell $i,j$, $S(i)$ is index set of the related cells, and $V$ is the volume of the cell.
	The interpolation operator $I_H^h$ is a bilinear interpolation operator. For the rectangular mesh as shown in Fig. \ref{BiIn}, let $(x_1,y_1), (x_1,y_2), (x_2,y_1), (x_2, y_2)$ be the four corner points of the current cell, $I_H^h$ is defined as
	\begin{equation}
		Q_{h,i} = I_H^h (\{Q_{H,j}, j \in S(i) \}) = \frac{\sum_{j \in S(i)}w_jQ_{H,j}}{\sum_{j \in S(i)}w_j},
	\end{equation}
	and
	\begin{equation}
		\begin{aligned}
			w_1 &= (x_2 - x_i)(y_2 - y_i), \\
			w_2 &= (x_i - x_1)(y_2 - y_i), \\
			w_3 &= (x_2 - x_i)(y_i - y_1), \\
			w_4 &= (x_i - x_1)(y_i - y_1).
		\end{aligned}
	\end{equation}
	Here, $(x_i, y_i)$ is the center coordinate of cell $i$.
	
	After these two operations are determined, the multi-grid method for solving Eq.~\eqref{nonlinear} and Eq.~\eqref{linear} can be constructed.
	For simplicity, the basic two-grid cycle method will be discussed, and the multiple V-cycle can be done in a recursive way.
	
	\subsection{Multi-grid method for macro-system}
	Since Eq.~\eqref{nonlinear} is a non-linear system, the full approximation scheme (FAS) \cite{brandt2011multigrid} will be used to solve this system. The algorithm for a two-grid cycle can be described as follows,
	\begin{itemize}
		\item Pre-smoothing: perform $\nu_1$ times LU-SGS to get an approximate solution $\vec{W}_{h,i}^*$ of Eq.~\eqref{nonlinear}, and then compute the residual by $\vec{r}_{h,i} = \vec{R}_{h,i}^n + \vec{A}\left(\vec{W}_{h,i}^n \right) - \vec{A}\left(\vec{W}_{h,i}^* \right)$.
		\item Restriction: restrict $\vec{W}_{h,i}^*$ and $\vec{r}_{h,i}$ to a coarse grid by $\vec{r}_{H,i} = I_{h}^{H}(\{ \vec{r}_{h,j}, j \in S(i)\})$ and $\vec{W}_{H,i}^* = I_{h}^{H}(\{ \vec{W}_{h,j}^*, j \in S(i)\})$.
		\item Smoothing: perform $\nu_2$ times LU-SGS to get an approximation solution $\vec{W}_{H,i}^{**}$ of the coarse grid problem $\vec{A}\left(\vec{W}_{H,i}\right) = \vec{r}_{H,i} + \vec{A}\left(\vec{W}_{H,i}^* \right)$.
		\item Correction: compute the error $\vec{e}_{H,i}$ in the coarse-grid by $e_{H,i} = \vec{W}_{H,i}^{**} - \vec{W}_{H,i}^{*}$, and then correct the approximation solution in the fine grid by $\vec{W}_{h,i}^* = \vec{W}_{h,i}^* + I_{H}^{h}(\{ \vec{e}_{H,j}, j \in S(i)\})$
		\item Post-smoothing: perform $\nu_3$ times LU-SGS to get the final solution $\vec{W}_{h,i}^{n+1}$ of Eq.~\eqref{nonlinear} with the initial guess $\vec{W}_{h,i}^*$.
	\end{itemize}
	\subsection{Multi-grid method for micro-system}
	For the linear system \eqref{linear}, the two-grid correction scheme can be summarized as,
	\begin{itemize}
		\item Pre-smoothing: perform $\gamma_1$ times SGS to get an approximate solution ${\Delta f}_{h,\{i,k\}}^*$ of Eq.~\eqref{linear}, and then compute the residual by $${r}_{h,\{i,k\}} =  (RHS)_{h,\{i,k\}} - \left( D_{h,\{i,k\}} \Delta f_{h,\{i,k\}}^{*} + \sum_{j \in {N(i)}}D_{h,\{j,k\}}\Delta f_{h,\{j,k\}}^{*}\right).$$
		\item Restriction: restrict ${\Delta f}_{h,\{i,k\}}^*$, ${r}_{h,\{i,k\}}$ and $D_{h,\{i,k\}}$ to a coarse grid by the restriction operator $I_{h}^{H}$.
		\item Smoothing: perform $\gamma_2$ times SGS to get a solution of the residual equation $$D_{H,\{i,k\}} \Delta e_{H,\{i,k\}} + \sum_{j \in {N(i)}}D_{H,\{j,k\}} \Delta e_{H,\{j,k\}} = r_{H,\{i,k\}}.$$
		\item Correction: correct the approximation solution in the fine grid by $\Delta f_{h,\{i,k\}}^{*}= \Delta f_{h,\{i,k\}}^{*} + I_{H}^{h}(\{ {e}_{H,\{i,k\}}, j \in S(i)\})$
		\item Post-smoothing: perform $\gamma_3$ times SGS to get the final solution $\Delta f_{h,\{i,k\}}^{n+1}$ of Eq.~\eqref{nonlinear} with the initial guess $\Delta f_{h,\{i,k\}}^{*}$.
	\end{itemize}
	The two-step IUGKS with single grid is just a reduction of the two-step IUGKS with multi-grid, where the pre-smoothing and post-smoothing steps are kept. Since the non-linear system may take more steps to get an accurate solution, the smoothing times $\nu_i$ for non-linear system are taken to be more than the smoothing times $\gamma_i$ for linear system. As demonstrated in \cite{trottenberg2000multigrid}, common choices are $\gamma_1 + \gamma_3 \leq 3$ in practice. In this paper, we set $\nu_1 = 10$, $\nu_2 = 10$ ,$\nu_3 = 5$ and $\gamma_1 = 2$, $\gamma_2 = 2$, $\gamma_3 = 1$.
	
	\section{Numerical validation}\label{example}
	
	In this section, several numerical examples will be given to validate the accuracy and efficiency of the IUGKS, including the extension to the full Boltzmann collision operator.
	For 2D cases, we will investigate the efficiency of three different implicit UGKS, which includes the original IUGKS \cite{zhu2016implicit}, the two-step IUGKS, and the two-step IUGKS with a V-cycle multi-grid solver.
	\subsection{Couette flow}
	The first case is the planar Couette flow for argon gas with $\omega = 0.81$. The full Boltzmann collision operator is considered in this case. The Knudsen number is defined by $\mathrm{Kn} = \mu_0 \sqrt{2\pi R T_0} / (2p_0 L)$, where $T_0$ is the reference temperature, $\rho_0$ is the reference density, $\mu_0$ is the viscosity, $p_0 = \rho_0 R T_0$ is the pressure, and $L$ is the characteristic length. The non-dimensional quantities are used in the numerical simulation. Therefore, without loss of generality, we set $T_0 = 1.0$, $\rho_0 = 1.0$ and $L = 1.0$. The velocities at the left and right walls are set as $U_L = 0.25$ and $U_R = -0.25$, respectively. The wall temperatures are fixed at $T_w = T_0 = 1.0$, and the diffusive boundary condition is adopted here.
	
\begin{figure}
		\centering
		\includegraphics[width=0.46\textwidth]{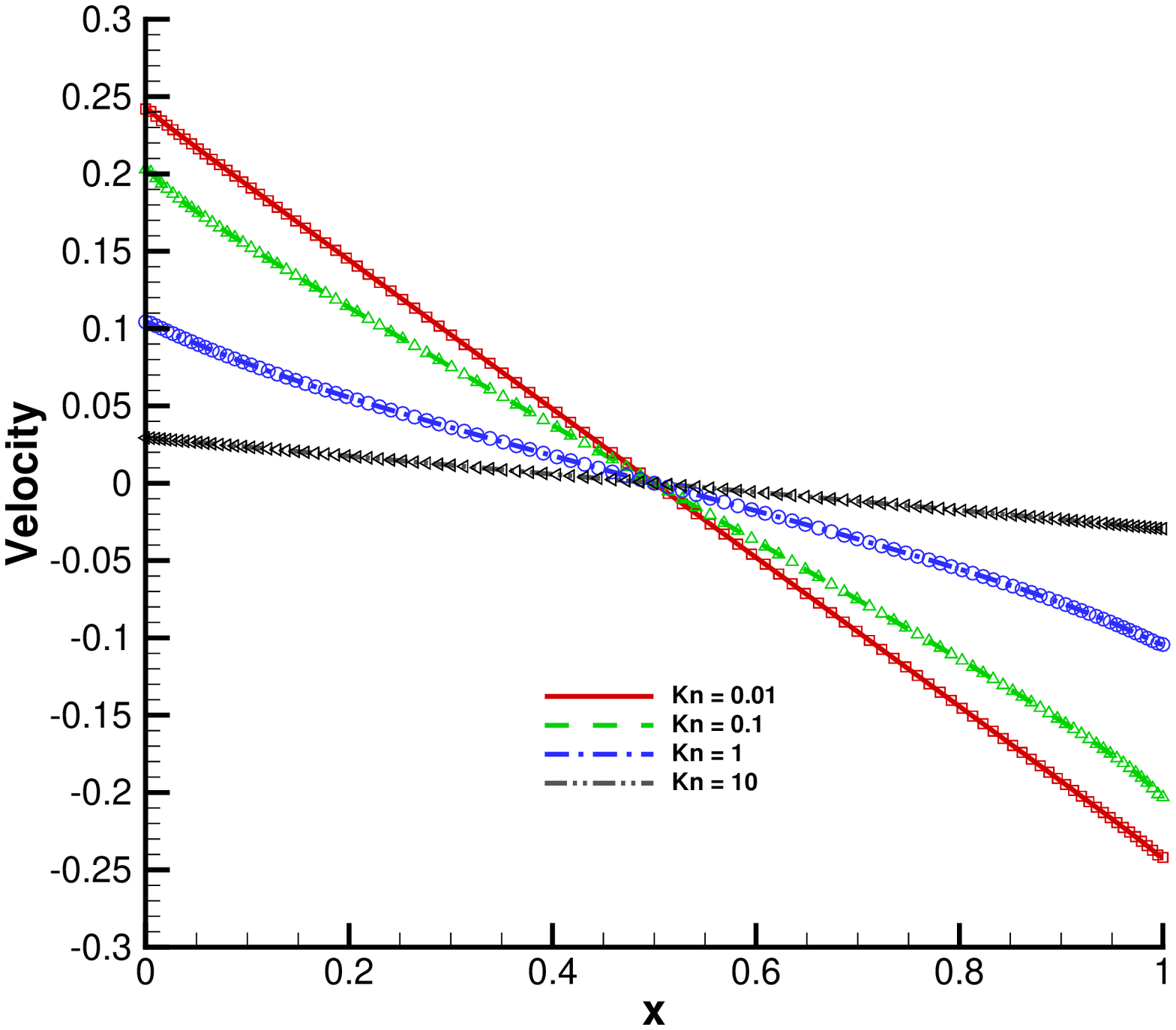}{(a)}
		\includegraphics[width=0.46\textwidth]{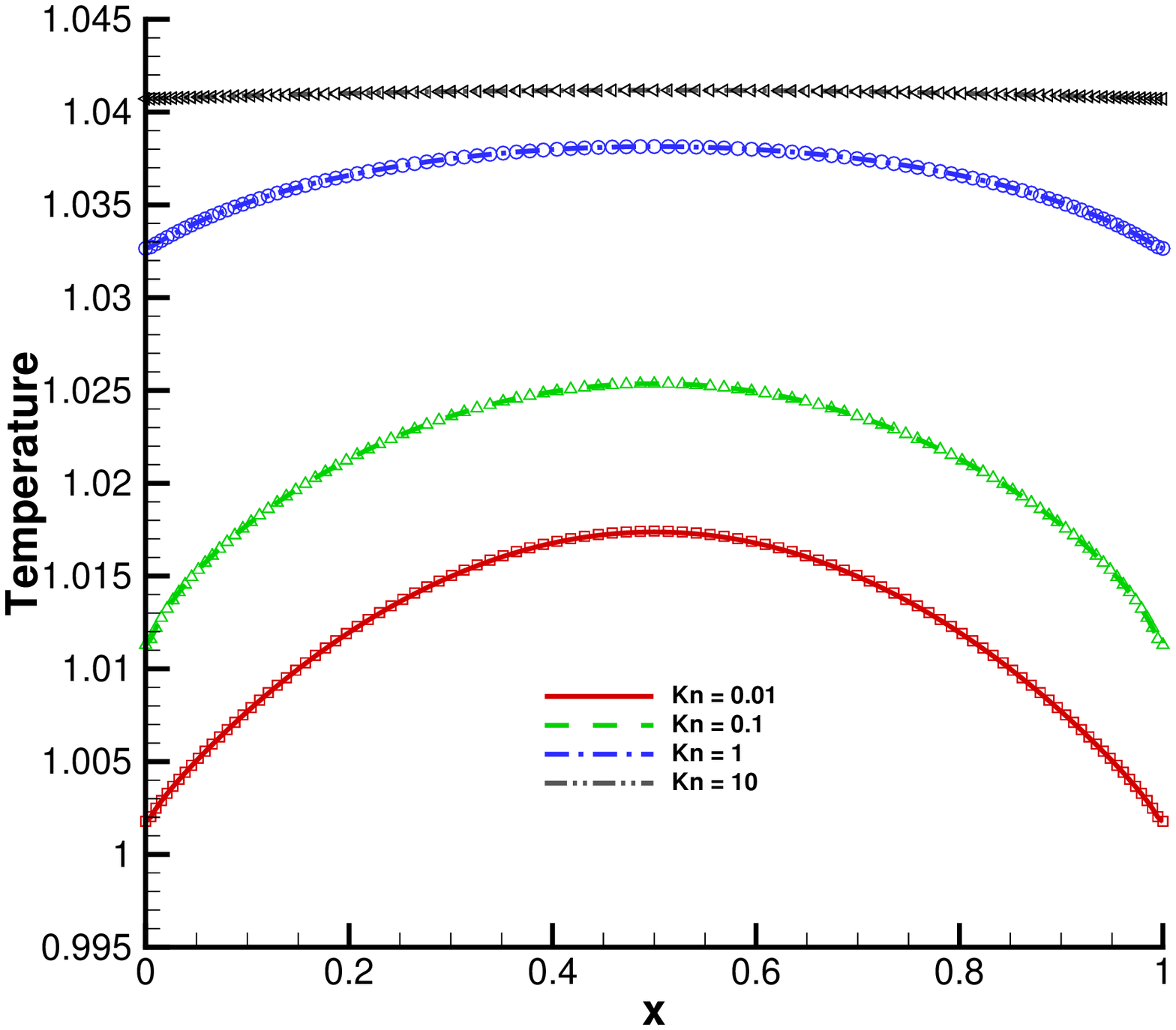}{(b)}
		\caption{(a) Velocity and (b) temperature profiles at different Knudsen numbers. The CIS solutions are shown in symbols, and the IUGKS solutions are shown in lines.}
		\label{couette}
	\end{figure}

\begin{figure}
		\centering
		\includegraphics[width=0.46\textwidth]{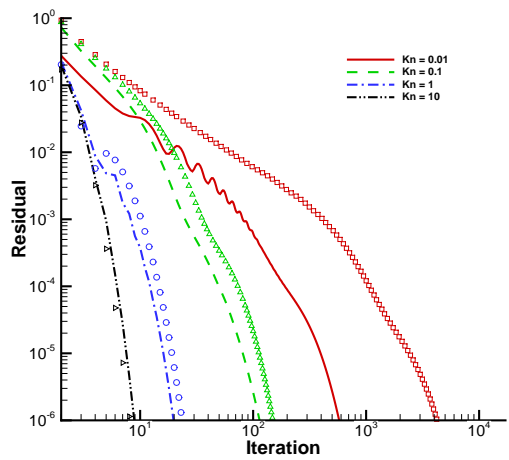}
		\caption{Convergence history of IUGKS and CIS in the planar Couette flow. The CIS results are shown in symbols, and the IUGKS results are shown in lines.}
		\label{Residualcouette}
	\end{figure}

	In the calculation, the spatial region is covered by $100$ unequally spaced cells with the minimum cell size $0.005$. The velocity space is truncated to $[-4.8,4.8]^3$, and there are $48$ uniform velocity mesh points in each direction. The cases with $Kn = 10,1,0.1,0.01$ are computed. The convergence criterion is set to be
	\begin{equation}
		\vec{E}^n = \frac{\sum_{i=1}^N \left|\vec{W}_i^n - \vec{W}_i^{n-1}\right| V_i}{\sum_i^N \left|\vec{W}_i^{n-1}\right|V_i} < 1 \times 10^{-6}.
	\end{equation}
	Here $N$ is the total number of the discrete cells in the computational domain. As shown in Fig. \ref{couette}, the velocity and temperature profiles computed by the two-step IUGKS match well with the reference solutions. The reference solutions are computed by the conventional iterative scheme (CIS) \cite{wu2013deterministic}, which reads as,
	\begin{equation}
		\nu(f^n)f^{n+1} + \vec{u}\cdot \nabla_{\vec{x}}f^{n+1} = Q^{+}(f^n,f^n),
	\end{equation}
	where $\nu$ is the collision frequency, and $Q^{+}$ is the gain term of the Boltzmann collision operator. Figure~\ref{Residualcouette} shows the convergence history of two-step IUGKS and CIS in the planar Couette flow. We can see that in the highly rarefied regime, both CIS and two-step IUGKS are efficient. However, as $\mathrm{Kn}$ decreases, the CIS needs many more steps to get the converged solutions than the two-step IUGKS.
	
	\subsection{Fourier flow}
	In the second case, we study the flow driven by the temperature gradient, and we still consider the full Boltzmann collision operator here. This case is similar to the planar Couette flow but with the stationary walls. The temperatures at the left and right walls are set as $T_L = 1.2$ and $T_R = 0.8$, respectively.
	
\begin{figure}
		\centering
		\includegraphics[width=0.46\textwidth]{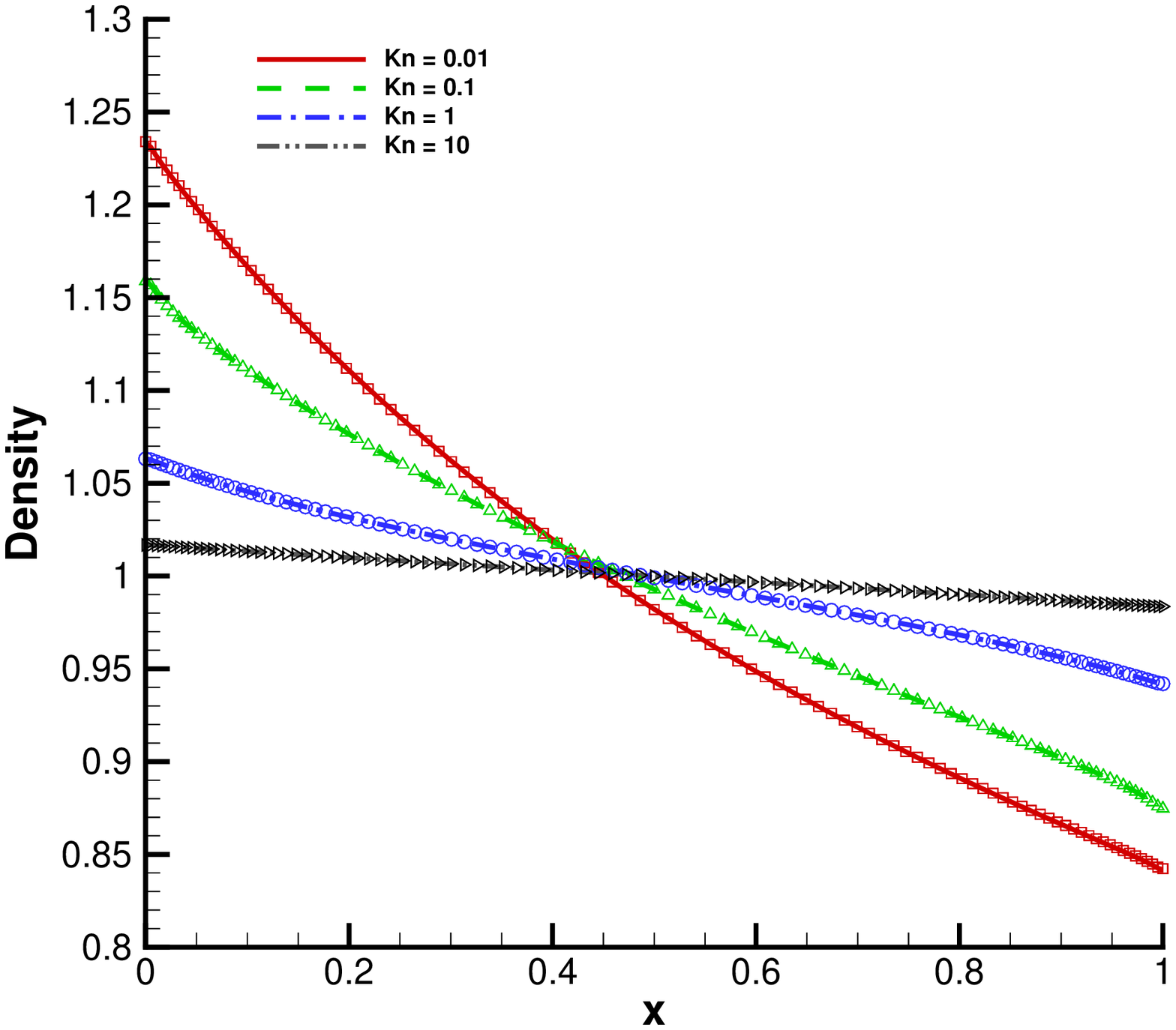}{(a)}
		\includegraphics[width=0.46\textwidth]{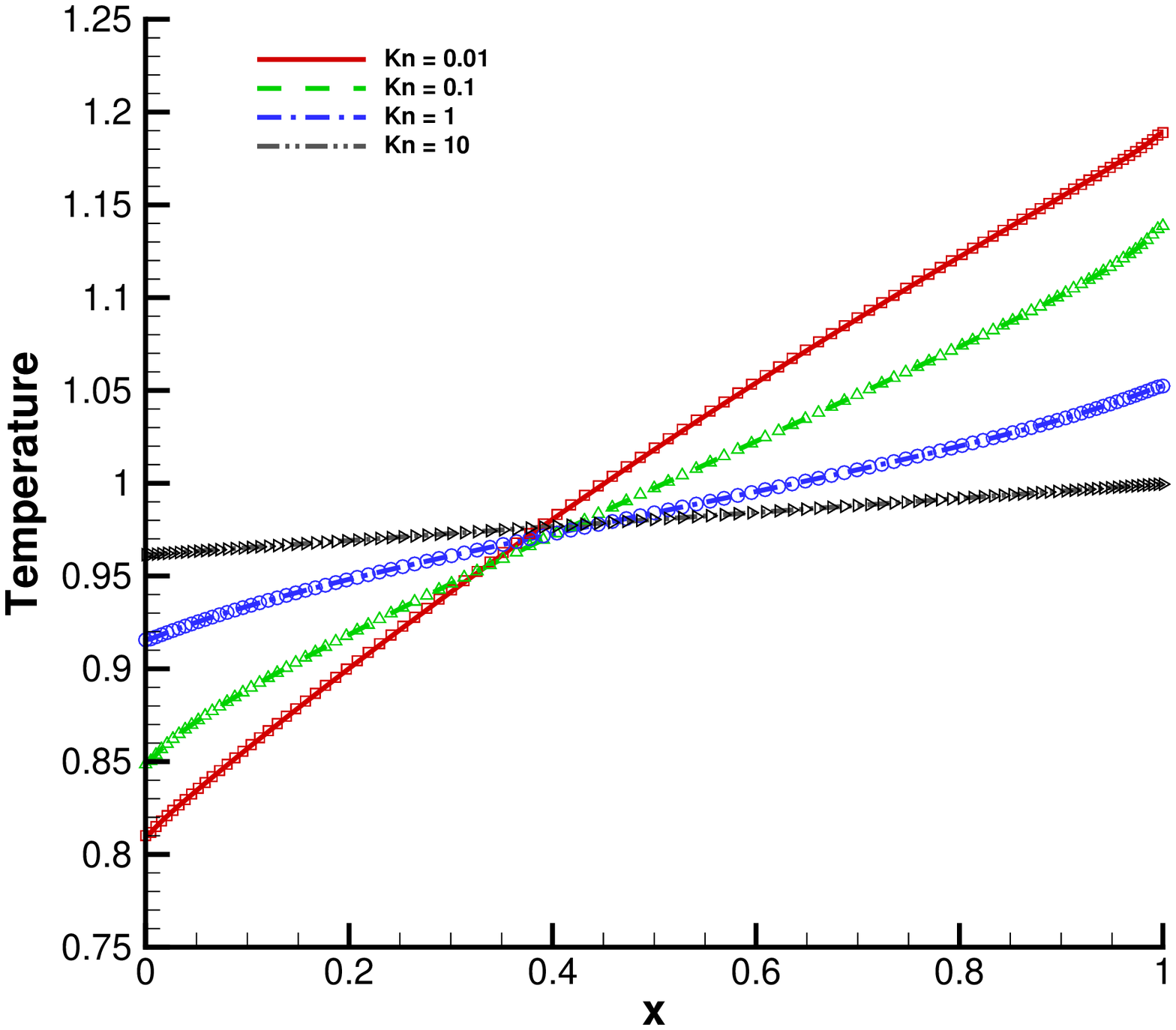}{(b)}
		\caption{(a) Density and (b) temperature profiles in different Knudsen number. The CIS solutions are shown in symbols, and the IUGKS solutions are shown in lines.}
		\label{fourier}
	\end{figure}

\begin{figure}
		\centering
		\includegraphics[width=0.46\textwidth]{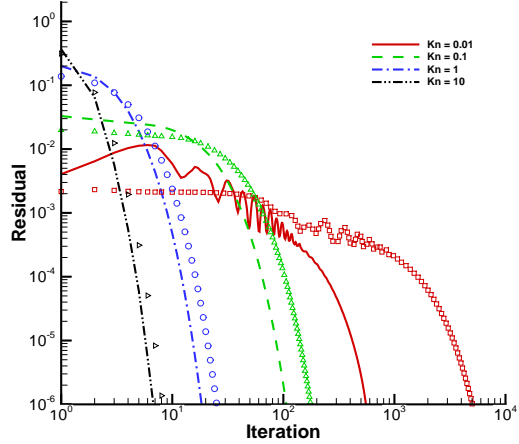}
		\caption{Convergence history of IUGKS and CIS in the planar Fourier flow. The CIS results are shown in symbols, and the IUGKS results are shown in lines.}
		\label{Residualfourier}
	\end{figure}

	The meshes in physical and velocity space and the convergence criterion are the same as the planar Couette flow.
	The cases at $Kn = 10,1,0.1,0.01$ have been tested. The density and temperature profiles in Fig. \ref{fourier} show that the two-step IUGKS present almost identical solutions as the CIS. However, as shown in Fig. \ref{Residualfourier}, the two-step IUGKS is more efficient than the CIS, especially when the Knudsen number becomes small.

	\subsection{Lid-driven cavity flow}
	The lid-driven cavity flow is studied at different Knudsen numbers. The collision model used in this case is the Shakhov model. The cavity has a fixed wall temperature with $T_w = 1.0$. The initial gas inside the cavity is argon gas with density $\rho_0 = 1.0$. The Knudsen number is defined by $\mathrm{Kn} = \mu_0 \sqrt{2\pi R T_0} / (2p_0 L)$, where $L = 1.0$ is the length of cavity side wall, and $\mu_0$ is evaluated by the variable hard sphere model with $\omega = 0.81$. The top wall has a velocity of $U_w = 0.148322$ in $x$ direction. A triple V-cycle multi-grid solver is used here.
	
\begin{figure}
		\centering
		\includegraphics[width=0.46\textwidth]{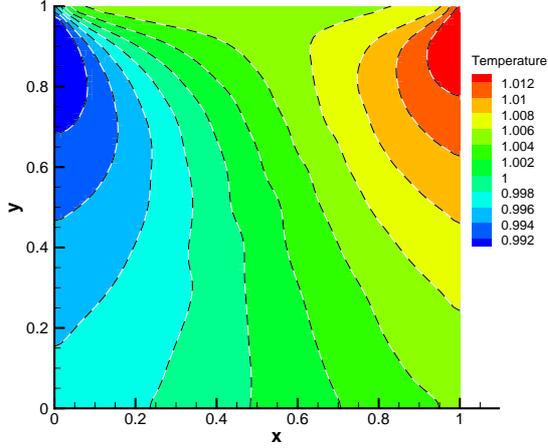}{(a)}
		\includegraphics[width=0.46\textwidth]{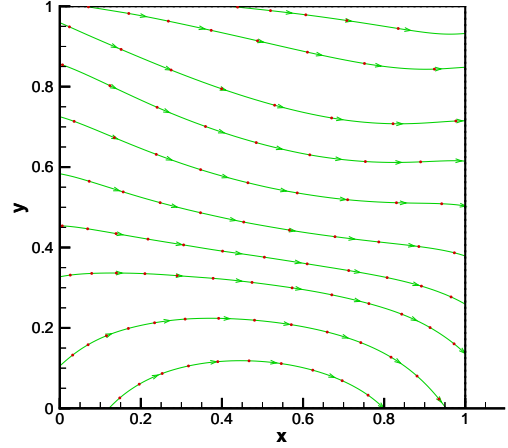}{(b)}
		\includegraphics[width=0.46\textwidth]{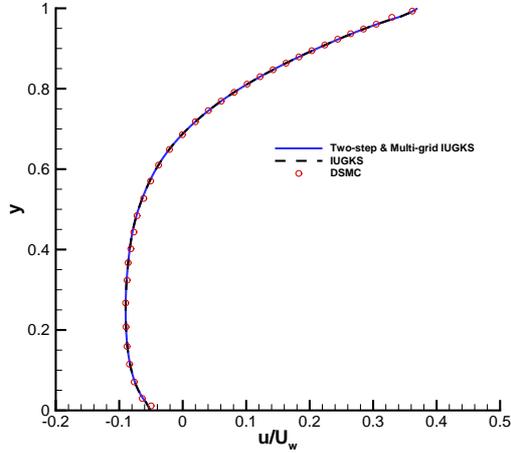}{(c)}
		\includegraphics[width=0.46\textwidth]{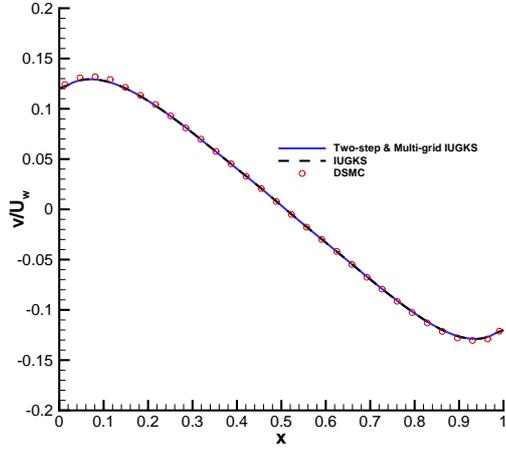}{(d)}
		\caption{Cavity flow at $\mathrm{Kn} = 10$. (a) Temperature. Background: two-step multi-grid IUGKS; Dashed line: IUGKS. (b) Heat flux. Line with arrowhead: two-step multi-grid IUGKS; Circle: IUGKS. (c) $U$-velocity along the central vertical line. (d) $V$-velocity along the central horizontal line.}
		\label{Kn10}
	\end{figure}

\begin{figure}
		\centering
		\includegraphics[width=0.46\textwidth]{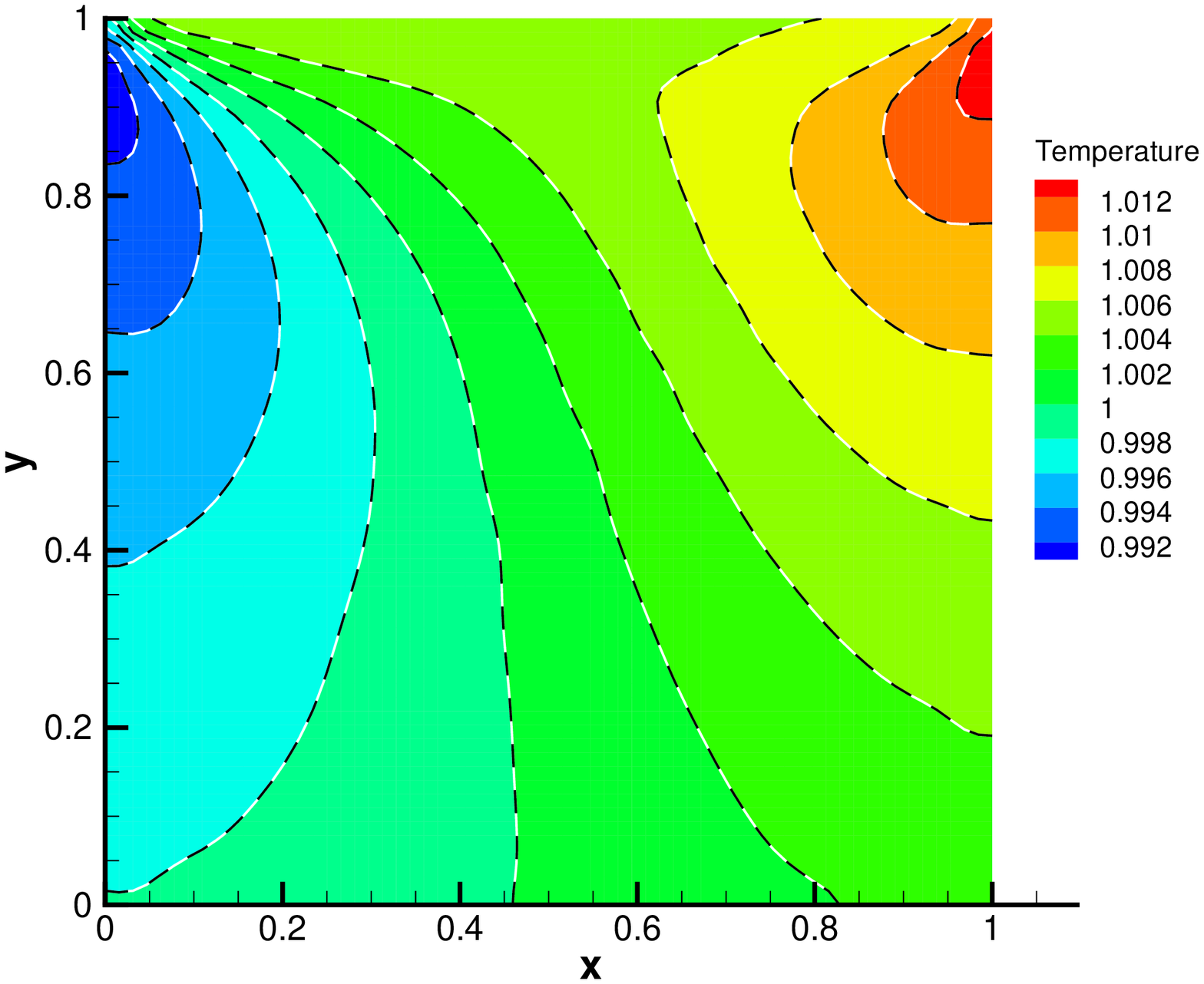}{(a)}
		\includegraphics[width=0.46\textwidth]{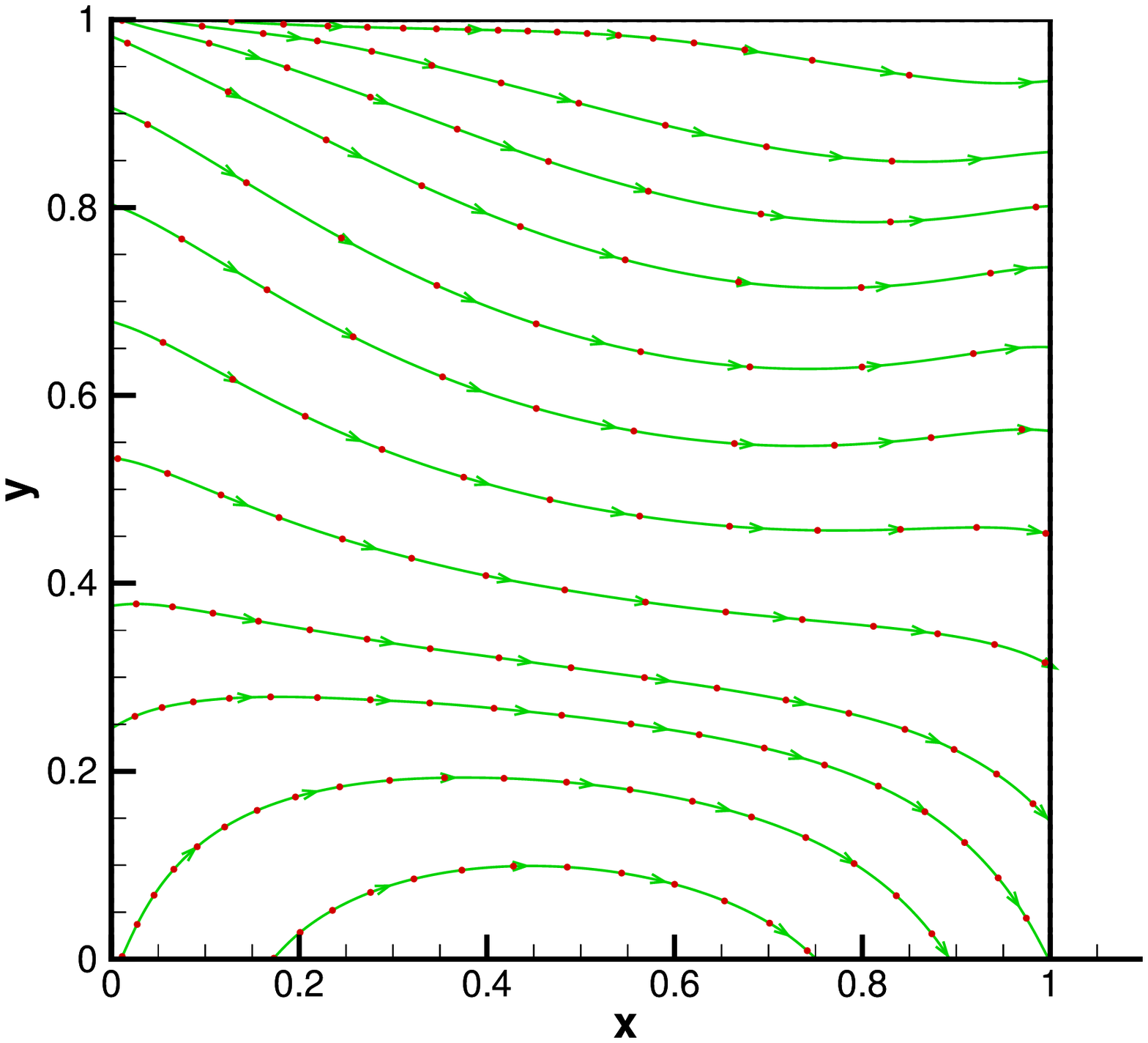}{(b)}
		\includegraphics[width=0.46\textwidth]{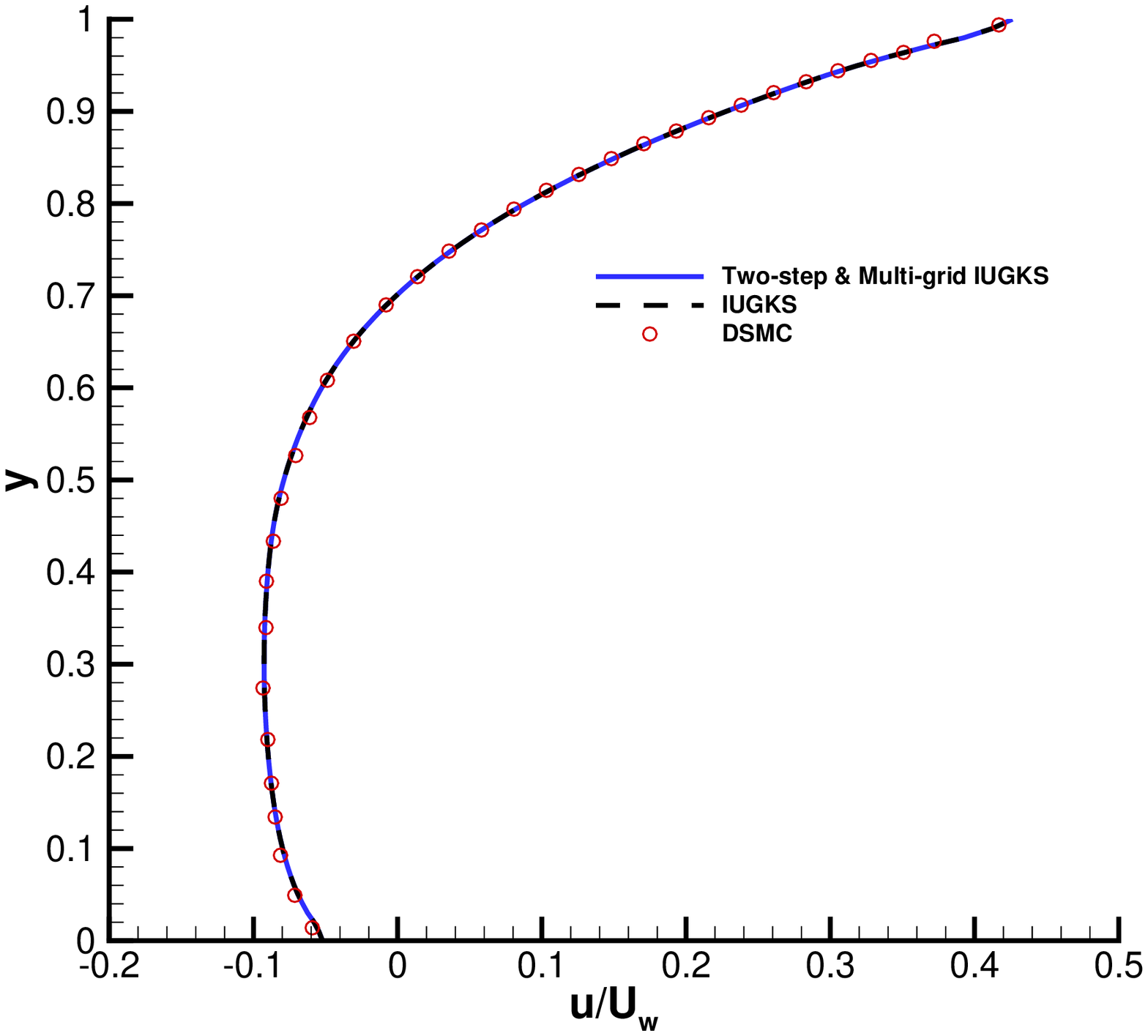}{(c)}
		\includegraphics[width=0.46\textwidth]{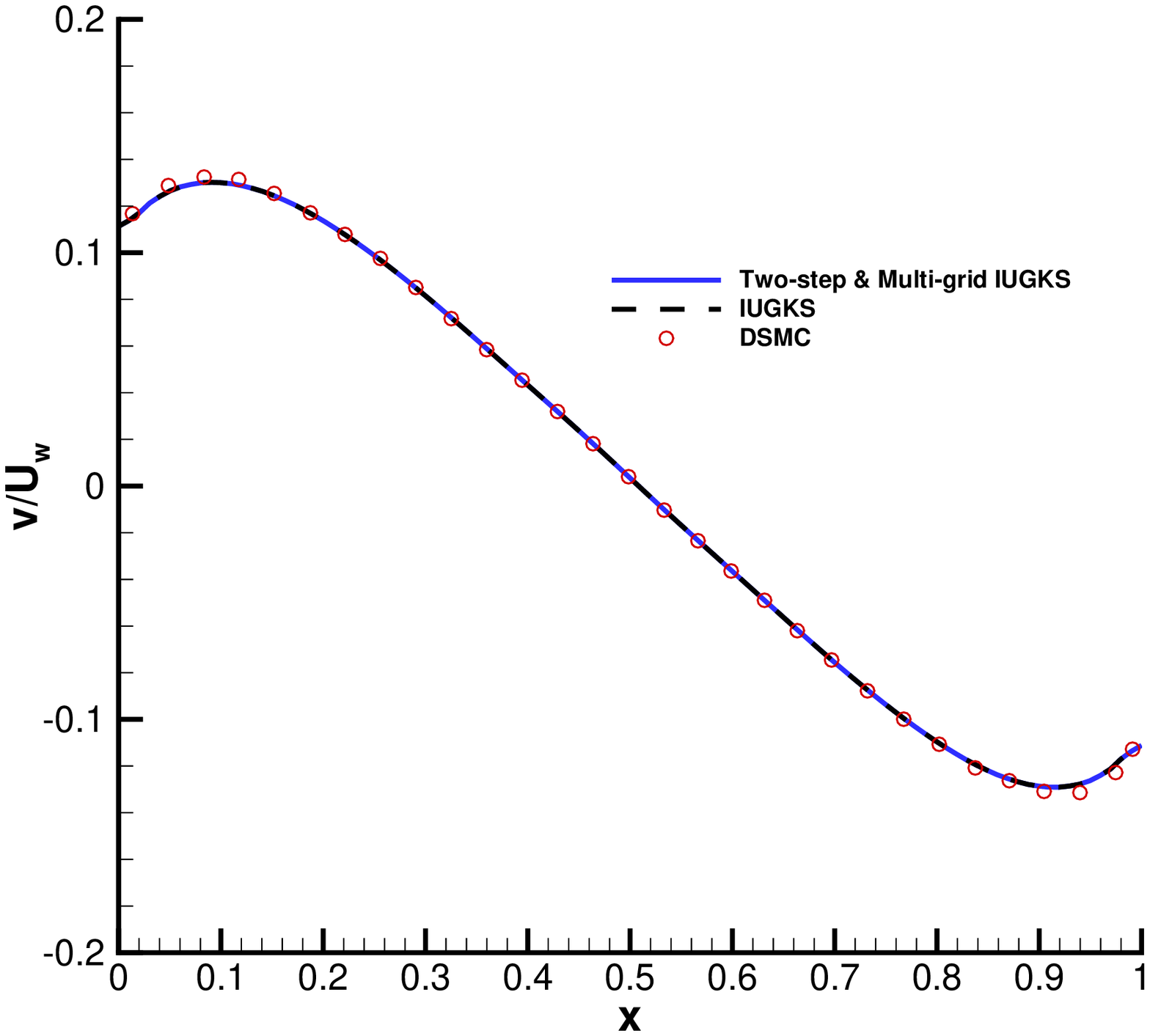}{(d)}
		\caption{Cavity flow at $\mathrm{Kn} = 1$. (a) Temperature. Background: two-step multi-grid IUGKS; Dashed line: IUGKS. (b) Heat flux. Line with arrowhead: two-step multi-grid IUGKS; Circle: IUGKS. (c) $U$-velocity along the central vertical line. (d) $V$-velocity along the central horizontal line.}
		\label{Kn1}
	\end{figure}

\begin{figure}
		\centering
		\includegraphics[width=0.46\textwidth]{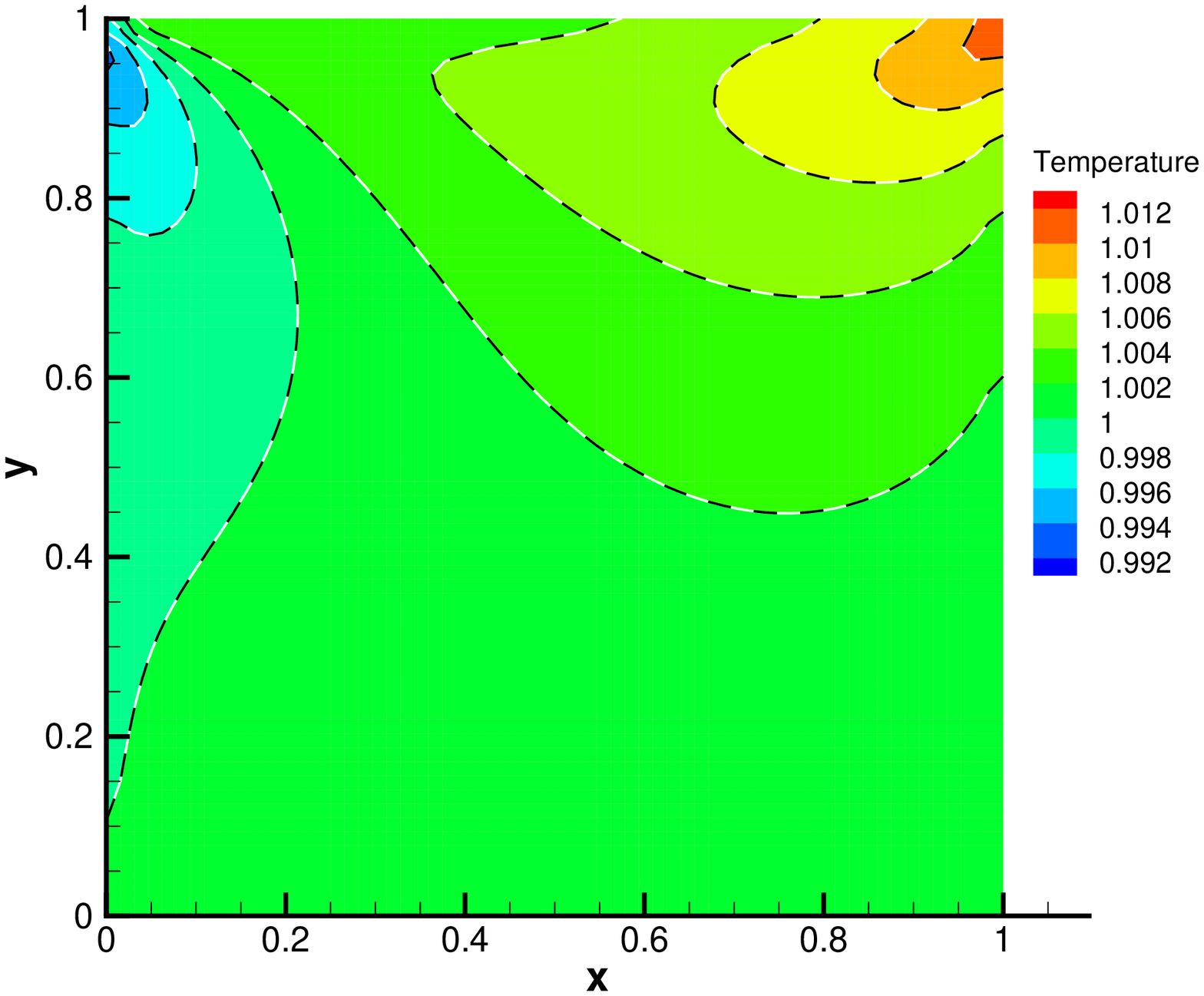}{(a)}
		\includegraphics[width=0.46\textwidth]{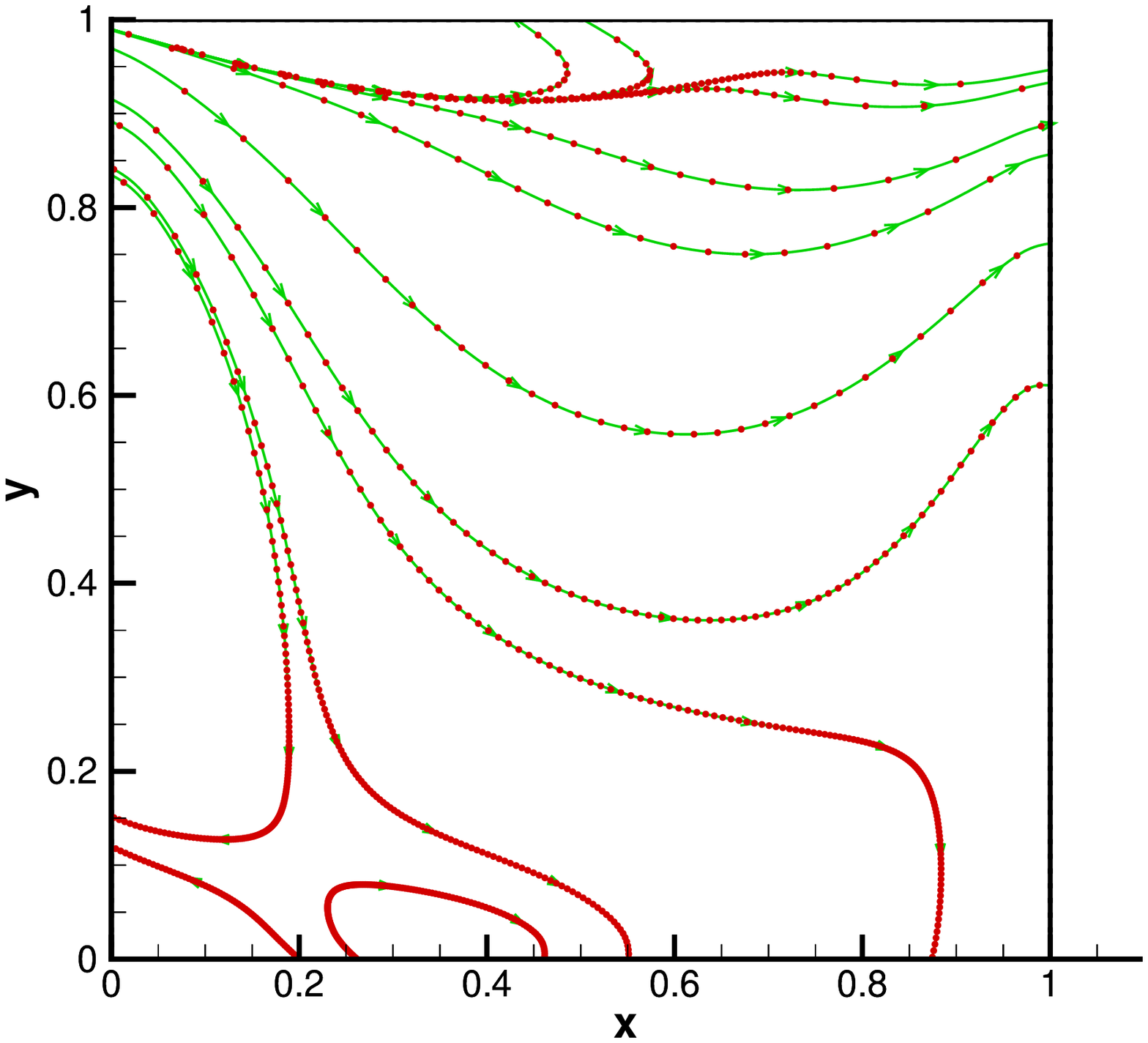}{(b)}
		\includegraphics[width=0.46\textwidth]{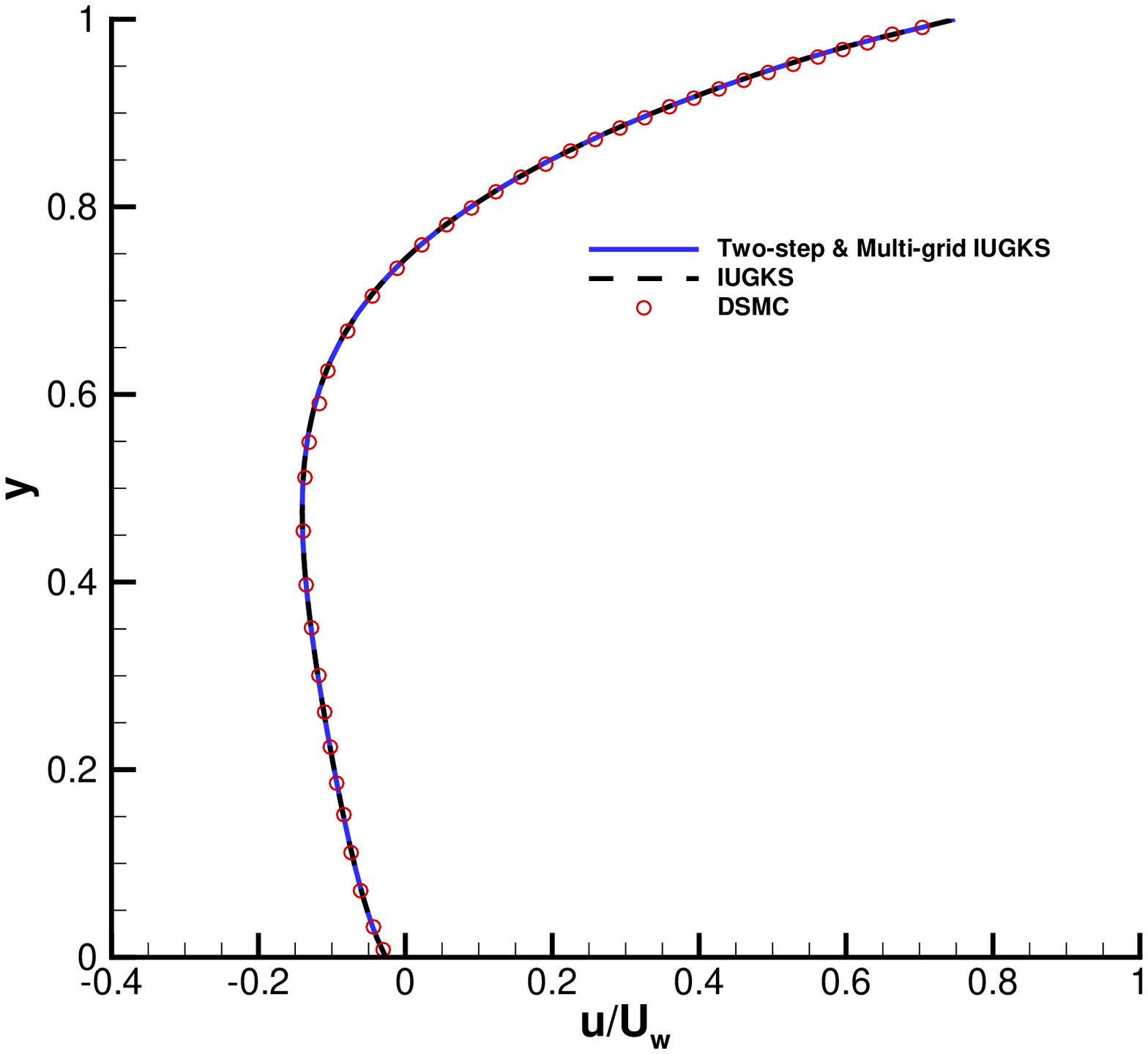}{(c)}
		\includegraphics[width=0.46\textwidth]{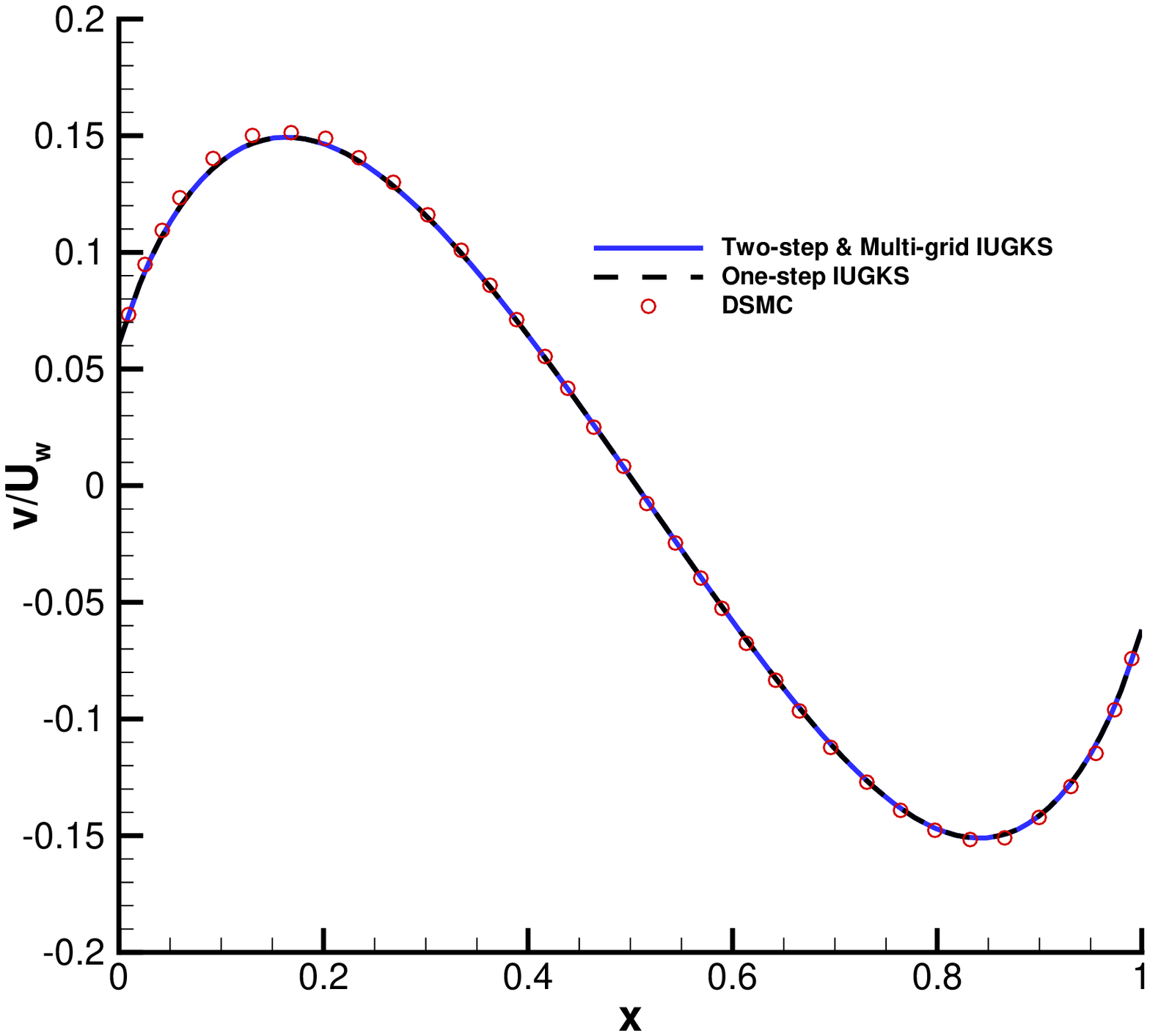}{(d)}
		\caption{Cavity flow at $\mathrm{Kn} = 0.075$. (a) Temperature. Background: two-step multi-grid IUGKS; Dashed line: IUGKS. (b) Heat flux. Line with arrowhead: two-step multi-grid IUGKS; Circle: IUGKS. (c) $U$-velocity along the central vertical line. (d) $V$-velocity along the central horizontal line.}
		\label{Kn0075}
	\end{figure}

	For the case $\mathrm{Kn} = 10, 1, 0.075$, the computational domain is discretized by a uniform mesh with $64 \times 64$ cells in the physical space. The particle velocity points are $80 \times 80$ and $60 \times 60$ for  $\mathrm{Kn} = 10$ and $\mathrm{Kn} = 1$, respectively. The trapezoidal integration is used to compute the moments of the distribution function in the two cases. While for the case of $\mathrm{Kn} = 0.075$, the Gauss-Hermite quadrature with $28 \times 28$ points in velocity space is adopted. The steady state is defined when the mean squared residuals of the conservative variables are reduced to a level being less than $1.0 \times 10^{-6}$, where the mean squared residuals are computed by
	\begin{equation}
		\vec{R}^n = \sqrt{\frac{\sum_{i=1}^{N} (\vec{R}_i^n)^2}{N}}.
	\end{equation}
	The results for $\mathrm{Kn} = 10, 1, 0.075$ are shown in Fig. \ref{Kn10}, \ref{Kn1} and \ref{Kn0075}, from which we can see that the results computed by the two-step IUGKS with multi-grid solver are identical to those computed by the original IUGKS.
	
\begin{figure}
		\centering
		\includegraphics[width=0.46\textwidth]{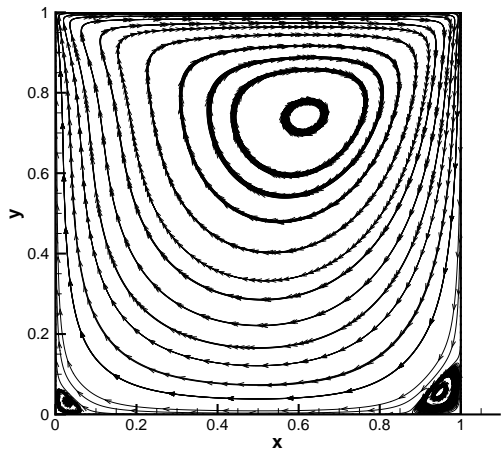}{(a)}
		\includegraphics[width=0.46\textwidth]{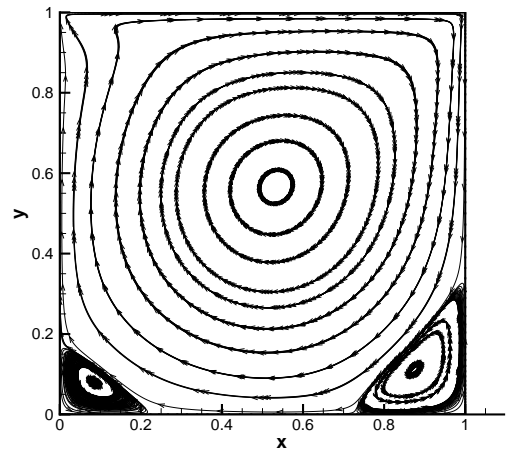}{(b)}
		\includegraphics[width=0.46\textwidth]{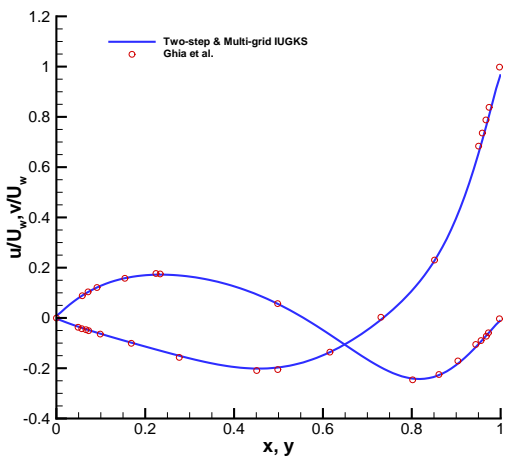}{(c)}
		\includegraphics[width=0.46\textwidth]{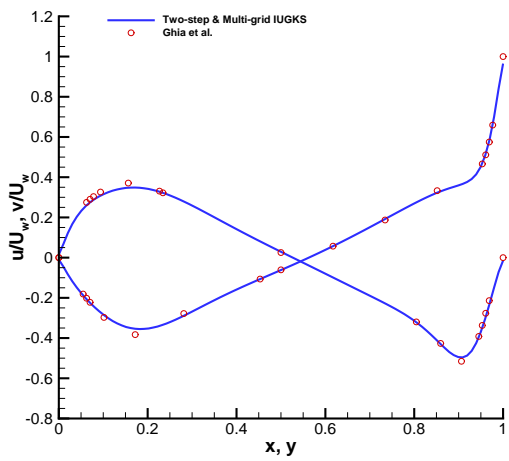}{(d)}
		\caption{Cavity flow at $\mathrm{Re} = 100$ (left) and $\mathrm{Re} = 1000$ (right). (a) and (b) Streamlines; (c) and (d) velocity along the central lines.}
		\label{Continnum}
	\end{figure}

\begin{figure}
		\centering
		\includegraphics[width=0.46\textwidth]{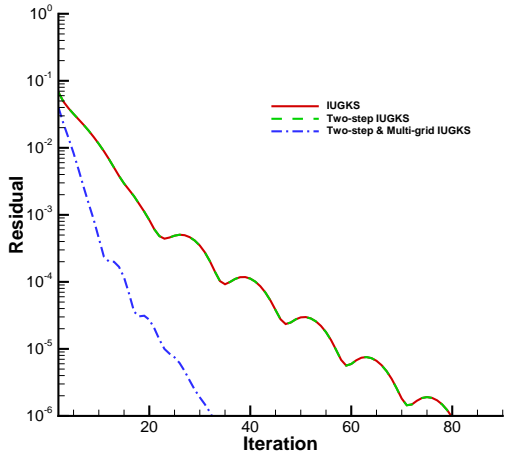}{(a)}
		\includegraphics[width=0.46\textwidth]{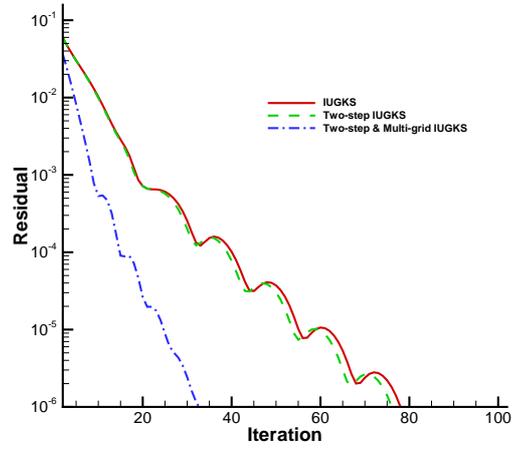}{(b)}
		\includegraphics[width=0.46\textwidth]{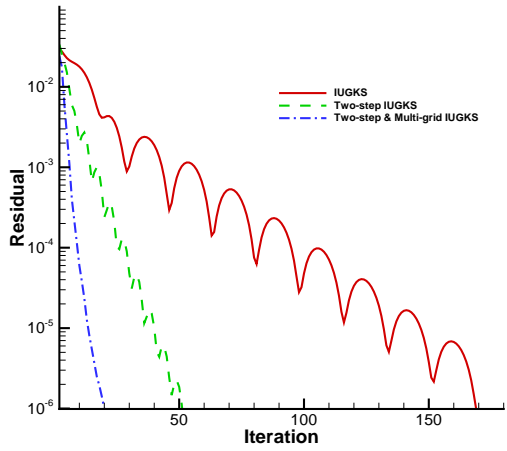}{(c)}
		\includegraphics[width=0.46\textwidth]{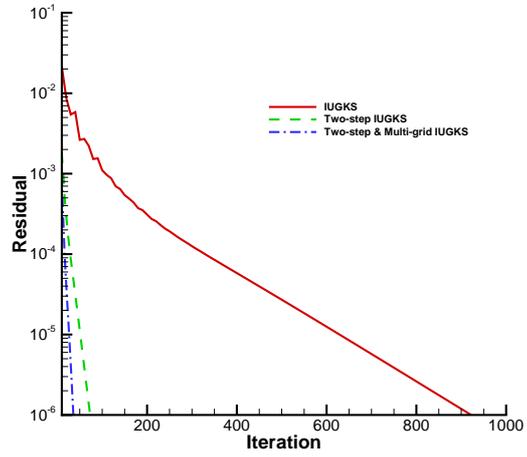}{(d)}
		\includegraphics[width=0.46\textwidth]{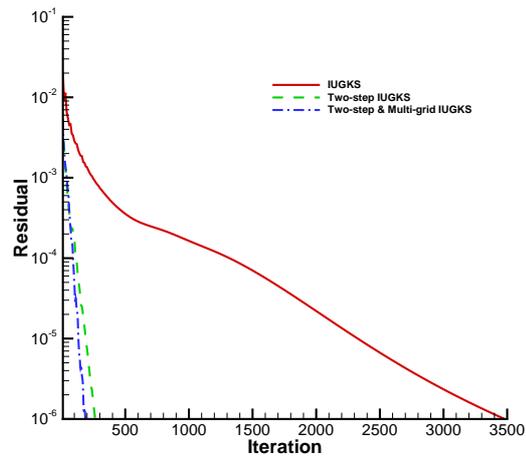}{(e)}
		\caption{Convergence history of different type of implicit UGKS in the cavity flow. (a) $\mathrm{Kn} = 10$; (b) $\mathrm{Kn} = 1$; (c) $\mathrm{Kn} = 0.075$; (d) $\mathrm{Re} = 100$; (e) $\mathrm{Re} = 1000$}
		\label{ContinnumResidual}
	\end{figure}

	Two continuum cases at $\mathrm{Re} = 100$ and $1000$ are also tested. The mesh is stretched to get a better resolution near the boundaries with $\Delta x_{min} = 0.004$. The particle velocity points are $28 \times 28$, and the Gauss-Hermite quadrature is used. The streamlines and the comparison with Ghia's data \cite{ghia1982high} are plotted in Fig. \ref{Continnum}. The comparison of the convergence history among different implicit UGKS is shown in Fig. \ref{ContinnumResidual} for all cases. The detailed computational time and acceleration rate in different flow regimes are listed in Table.\ref{cavityeff}. We can see that for the rarefied cases at $\mathrm{Kn} = 10$ and $1$, the convergence history of the original IUGKS and two-step IUGKS are similar, and the multi-grid solver can improve the efficiency of the IUGKS. As the decrease of the Knudsen number, the two-step acceleration begins to take effect. Especially for the continuum case, the improvement of the efficiency in comparison with the original IUGKS is satisfactory.

	\begin{table}
		\caption{Efficiency of different implicit UGKS for cavity flow}
		\begin{tabular}{llllllllll}
			\hline
			\multirow{2}{*}{State} & \multicolumn{2}{l}{IUGKS} &  & \multicolumn{2}{l}{Two-step IUGKS} &  & \multicolumn{3}{l}{Two-step $\&$ Multi-grid IUGKS} \\ \cline{2-10}
			& Steps & Time(min) &  & Steps & Time(min) &  & Steps & Time(min) & Rate \\ \hline
			$\mathrm{Kn} = 10$    & 80    & 13.8      &  & 80    & 13.8      &  & 33    & 7.7       & 1.8  \\
			$\mathrm{Kn} = 1$     & 79    & 7.7       &  & 77    & 7.6       &  & 33    & 4.4      & 1.75  \\
			$\mathrm{Kn} = 0.075$ & 169   & 3.6         &  & 51   & 1.2       &  & 20    & 0.6       & 6.0  \\
			$\mathrm{Re} = 100$   & 923   & 19.5      &  & 74   & 1.7       &  & 36    & 1.1       & 17.7  \\
			$\mathrm{Re} = 1000$ & 3480  & 73.5      &  & 263  & 6.0      &  & 187   & 5.6      & 13.1  \\ \hline
		\end{tabular}
		\label{cavityeff}
	\end{table}
	
	\subsection{Hypersonic flow past a square cylinder}
	The last example is the hypersonic flow passing over a square cylinder at Knudsen numbers $1$ and $0.1$. The Knudsen number is defined by
	\begin{equation}
		\mathrm{Kn} = \frac{(5 - 2\omega)(7-2\omega)\mu_{\infty} \sqrt{2R T_{\infty}}}{15\sqrt{\pi}p_{\infty} L}.
	\end{equation}
	Here, $L = 1$ is the diameter of the square cylinder and $\omega = 0.81$.
	The free streaming argon gas has a speed of $\mathrm{Ma} = 5$ and a temperature $T_{\infty} = 1$. The temperature of the solid surface of the square cylinder is set to be $T_w = T_{\infty}$, and the diffusive boundary condition is adopted in the calculations. Due to the symmetry, the solution on the half of the physical domain will be calculated.
	
	The computational domain is discretized unequally with $\Delta x_{min} = 0.01$. For the velocity space, $101 \time 101$ uniform discrete velocity points are used. The Newton-Cotes rule is applied here to compute the numerical integration. A two-layer V-cycle multi-grid solver is used here. In this hypersonic case, the convergence criterion is set as $\vec{R}^n < 1 \times 10^{-4}$.
	
\begin{figure}
		\centering
		\includegraphics[width=0.46\textwidth]{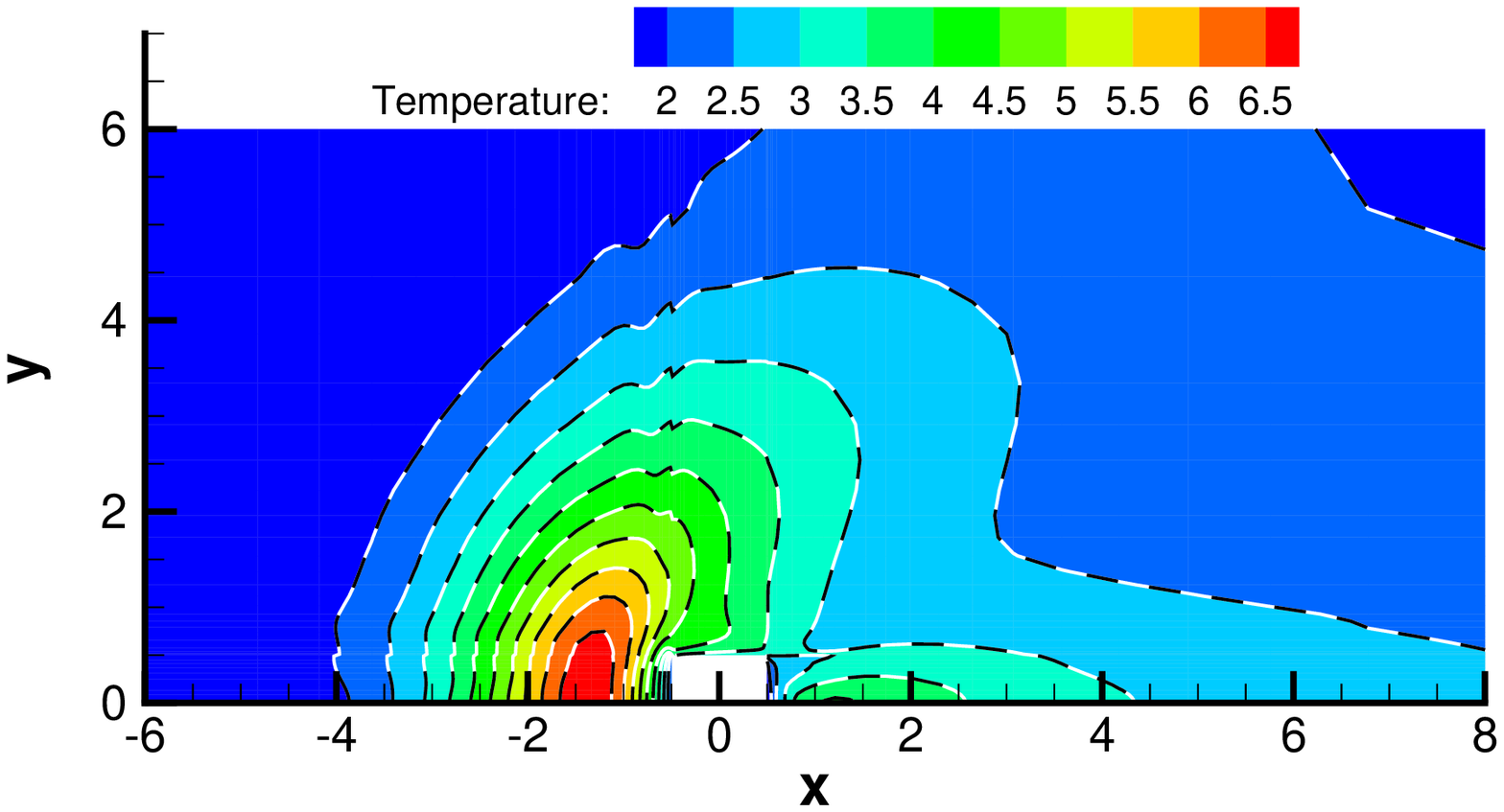}{(a)}
		\includegraphics[width=0.46\textwidth]{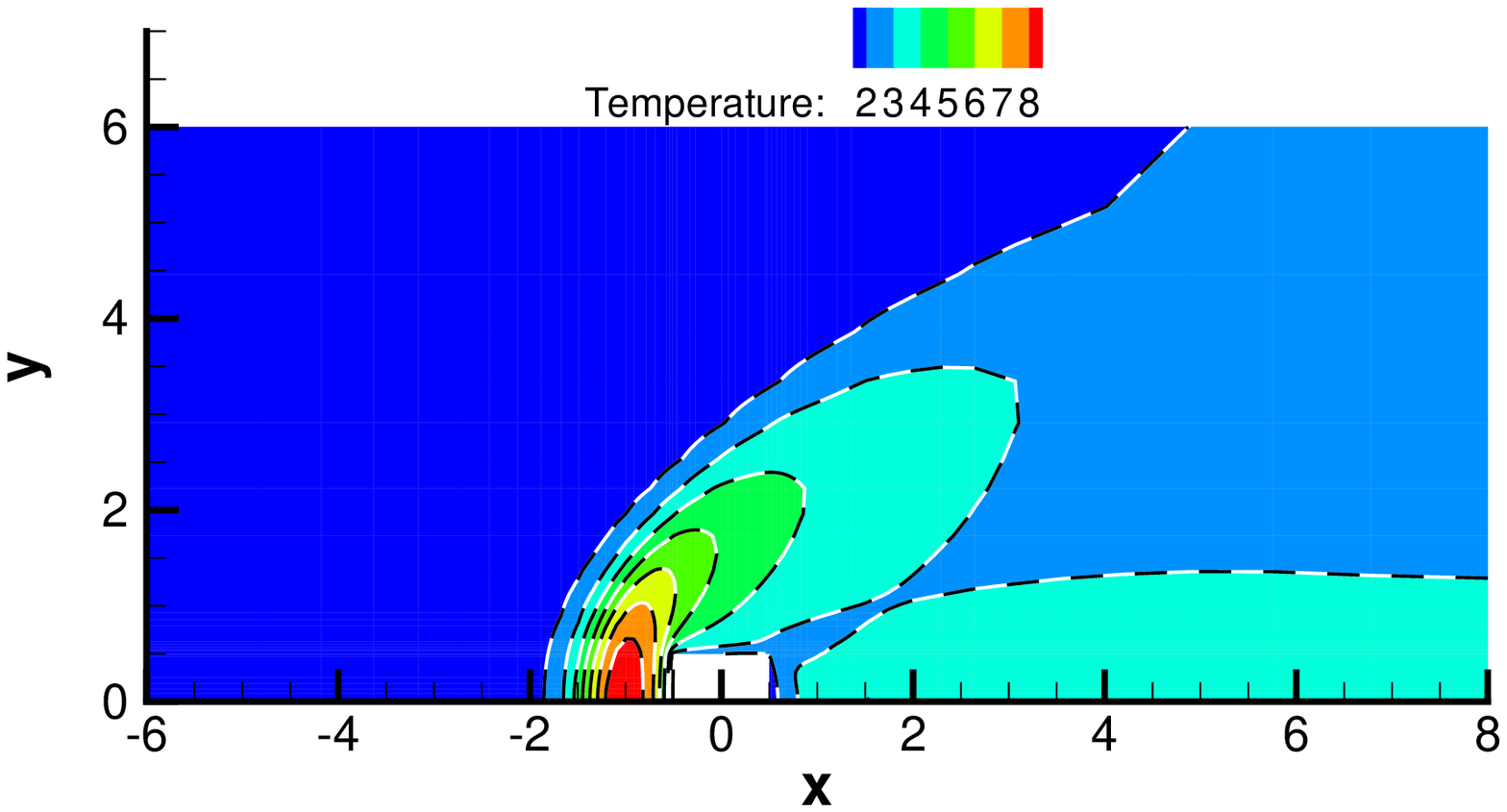}{(b)}
		\includegraphics[width=0.46\textwidth]{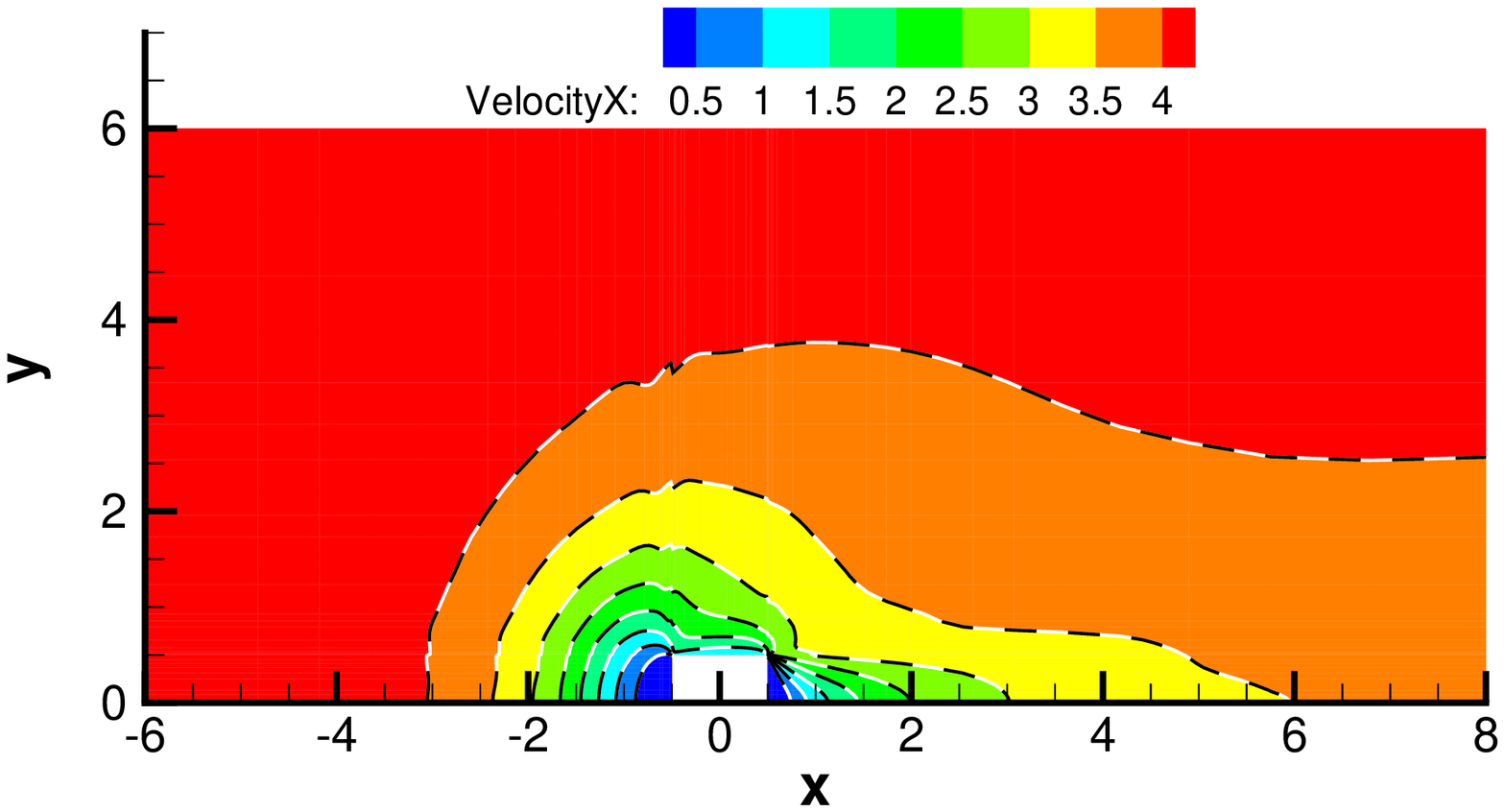}{(c)}
		\includegraphics[width=0.46\textwidth]{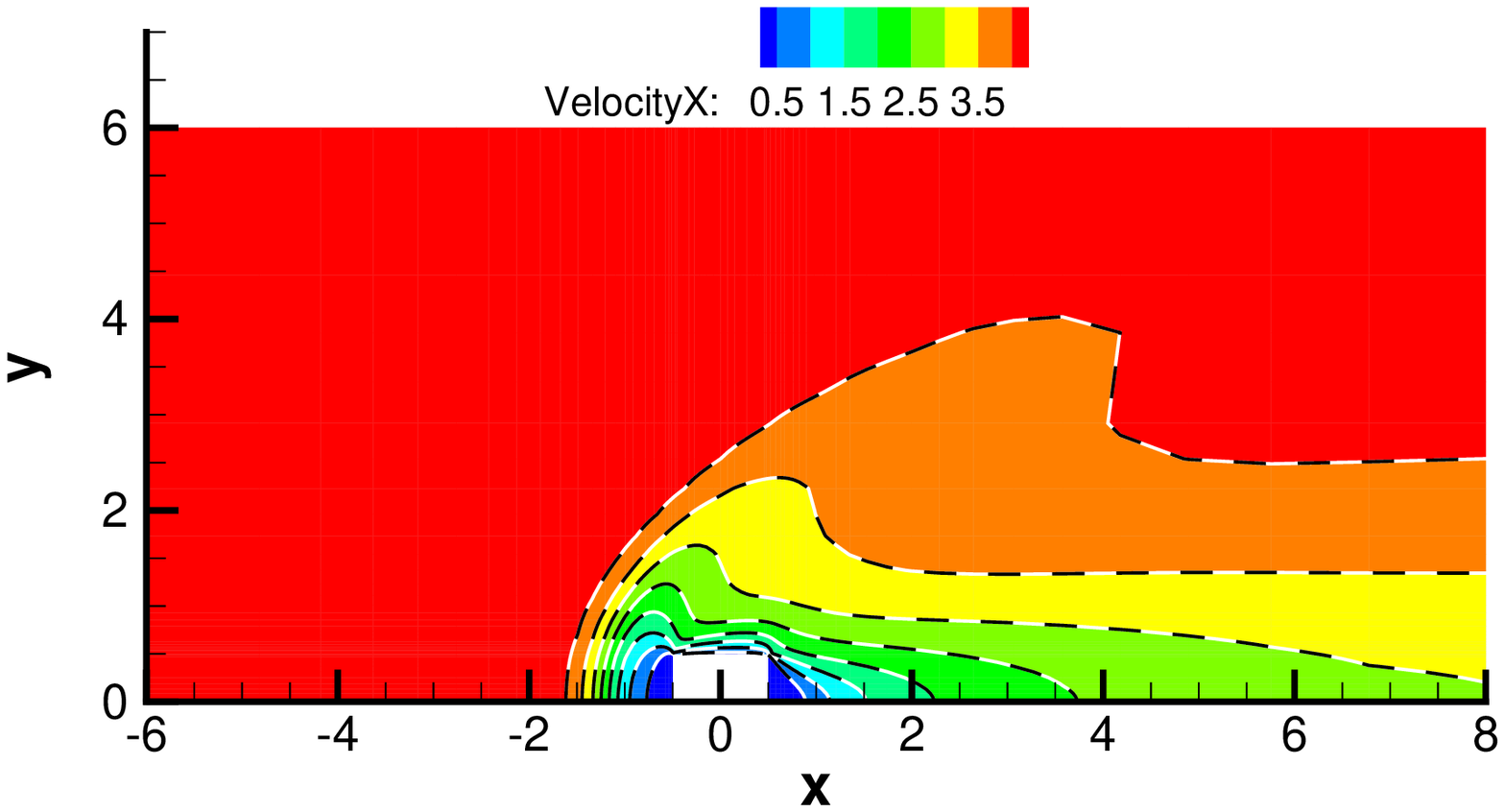}{(d)}
		\includegraphics[width=0.46\textwidth]{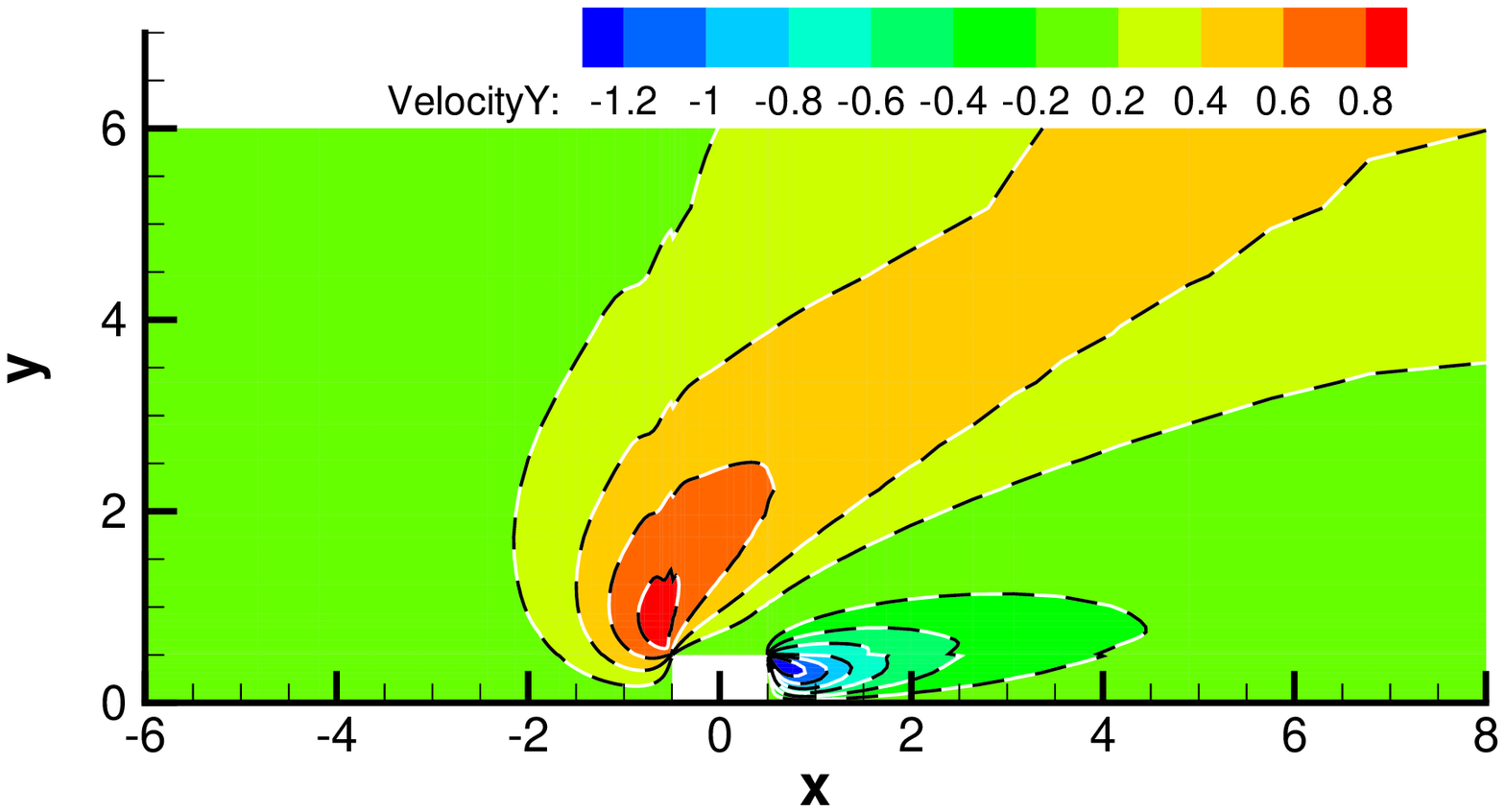}{(e)}
		\includegraphics[width=0.46\textwidth]{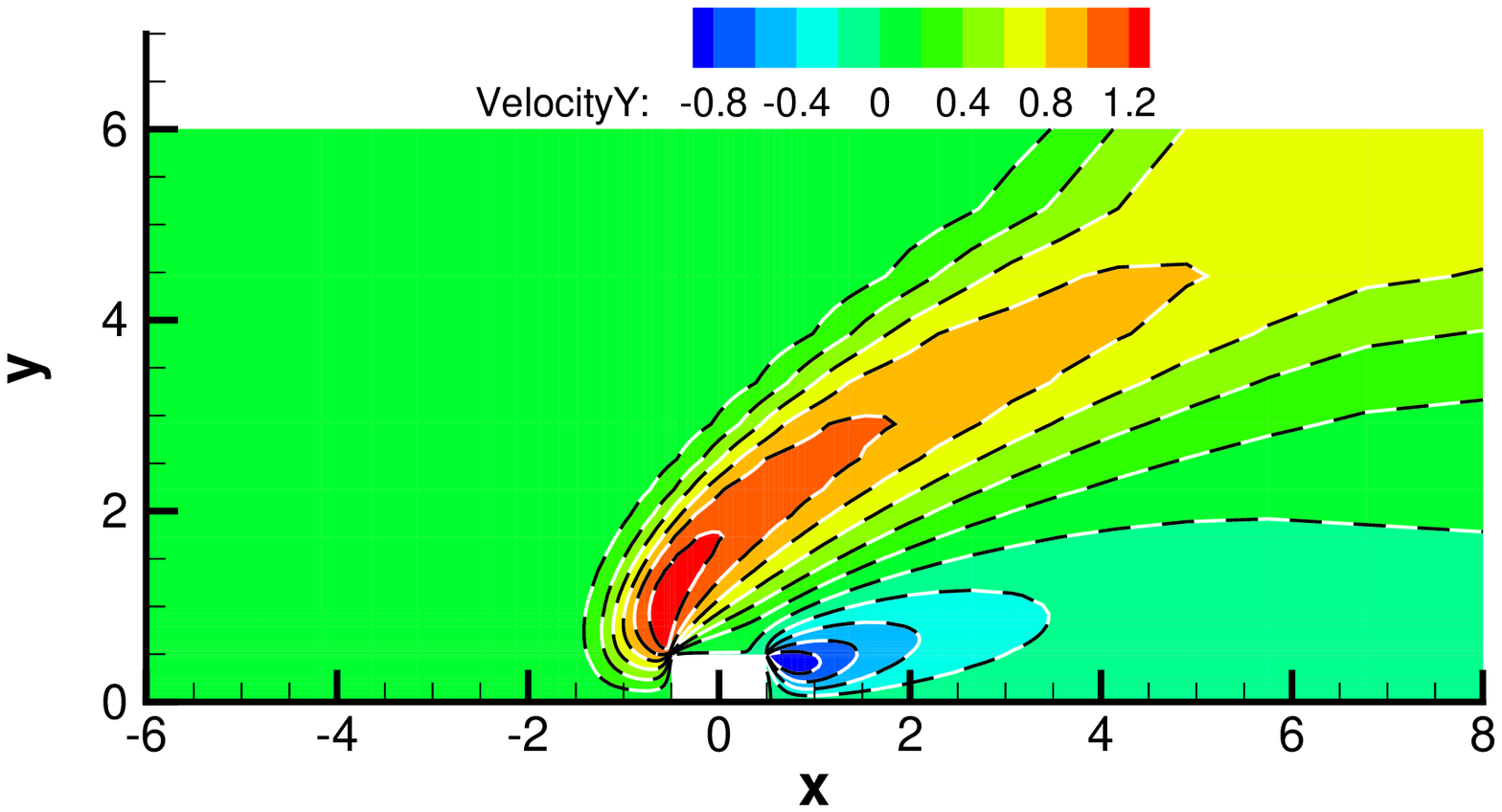}{(f)}
		\caption{Hypersonic flow around a square cylinder at $\mathrm{Kn} = 1$ (left) and $\mathrm{Kn} = 0.1$ (right). (a) and (b) Temperature; (c) and (d) $X$-velocity; (e) and (f) $Y$-velocity. Background: two-step multi-grid IUGKS; Dashed line: IUGKS.}
		\label{Squre}
	\end{figure}

\begin{figure}
		\centering
		\includegraphics[width=0.46\textwidth]{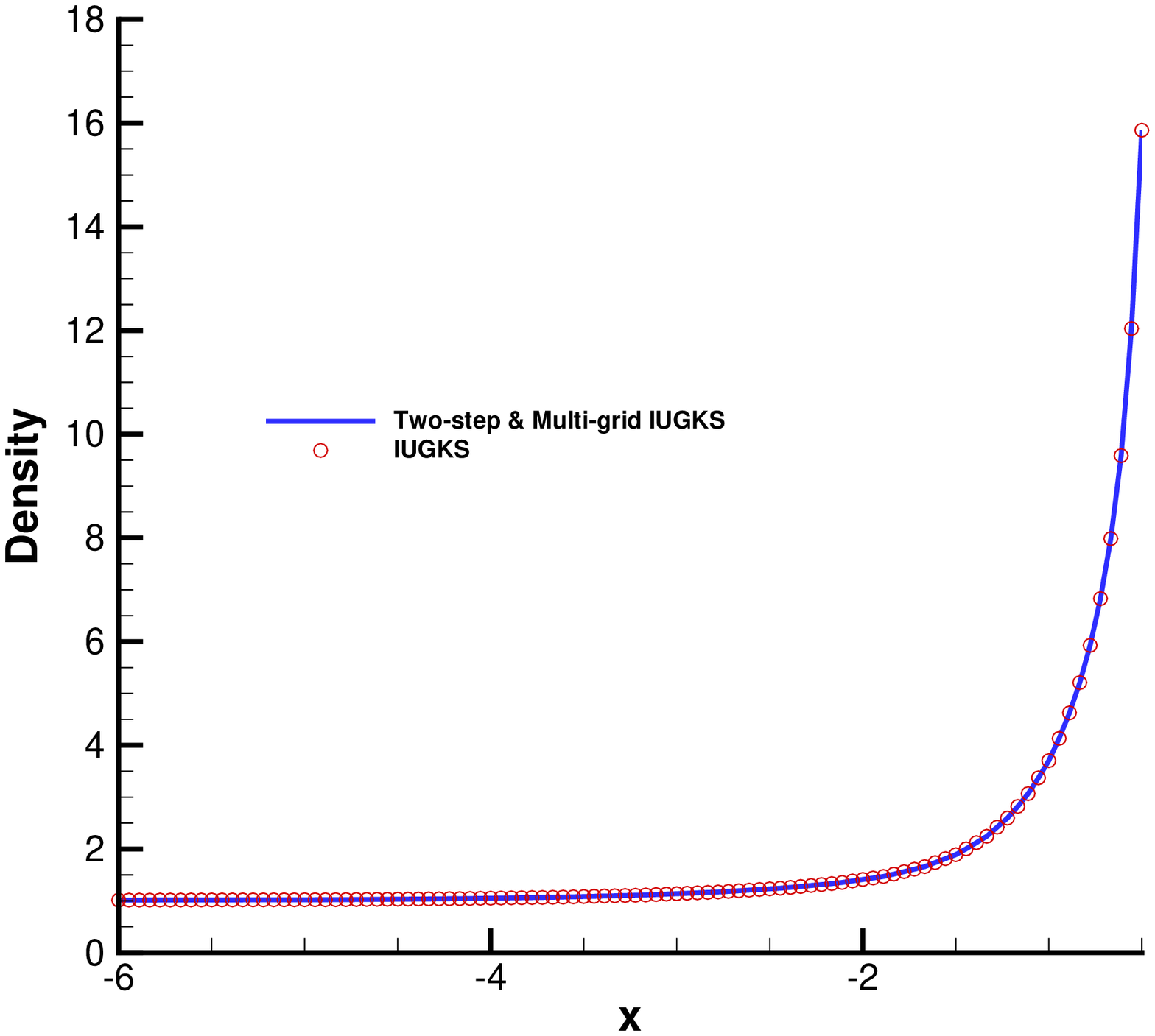}{(a)}
		\includegraphics[width=0.46\textwidth]{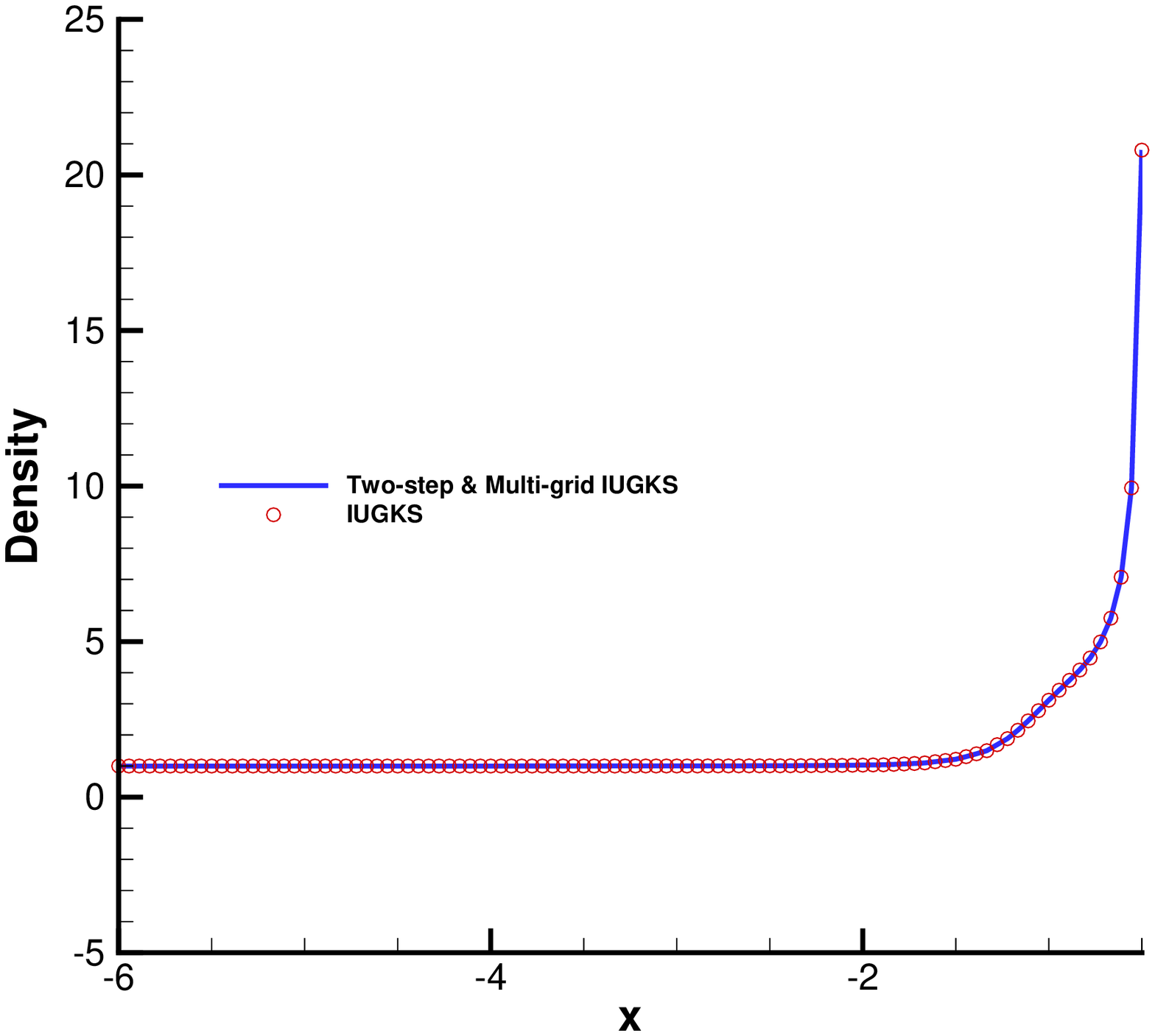}{(b)}
		\includegraphics[width=0.46\textwidth]{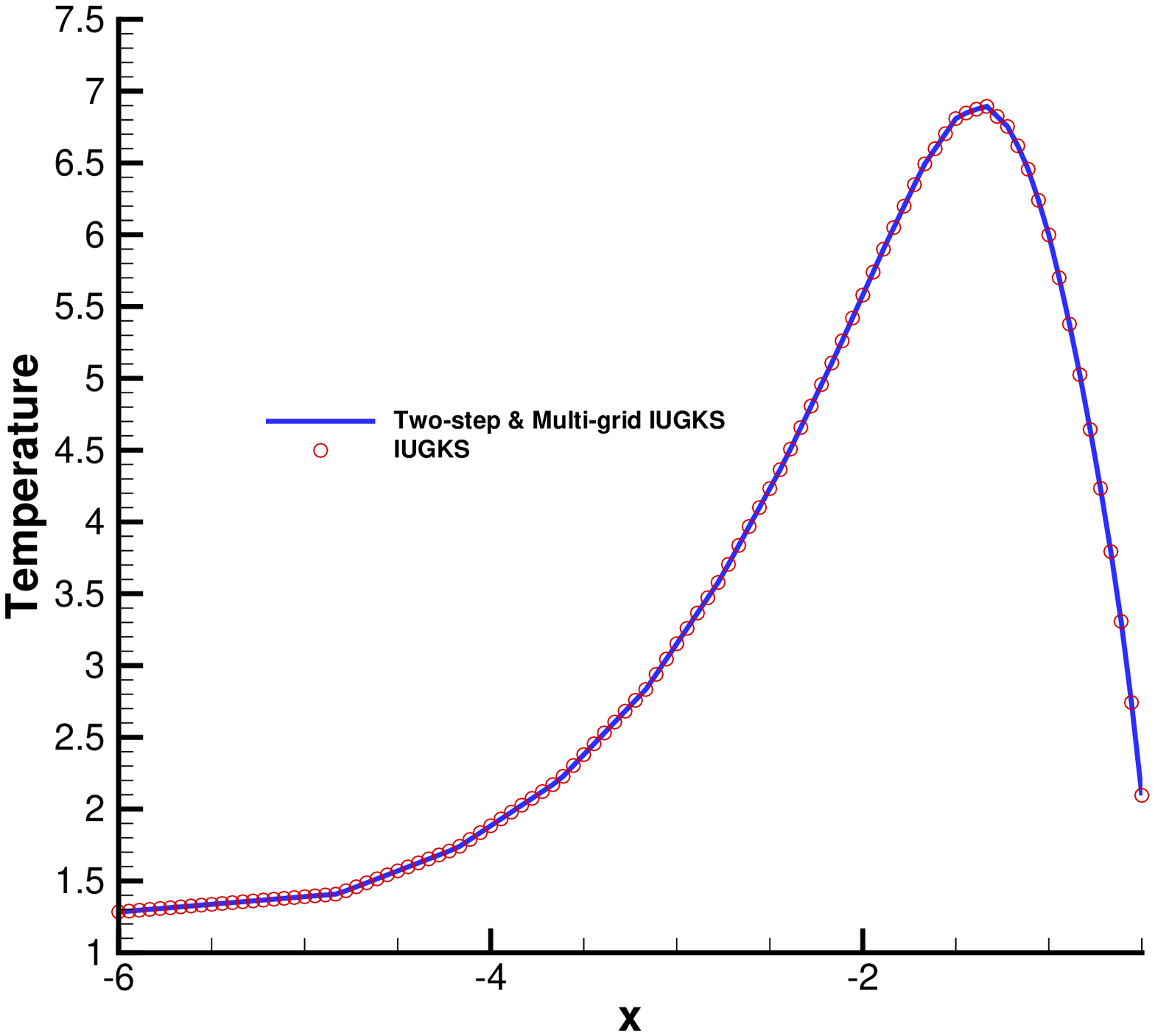}{(c)}
		\includegraphics[width=0.46\textwidth]{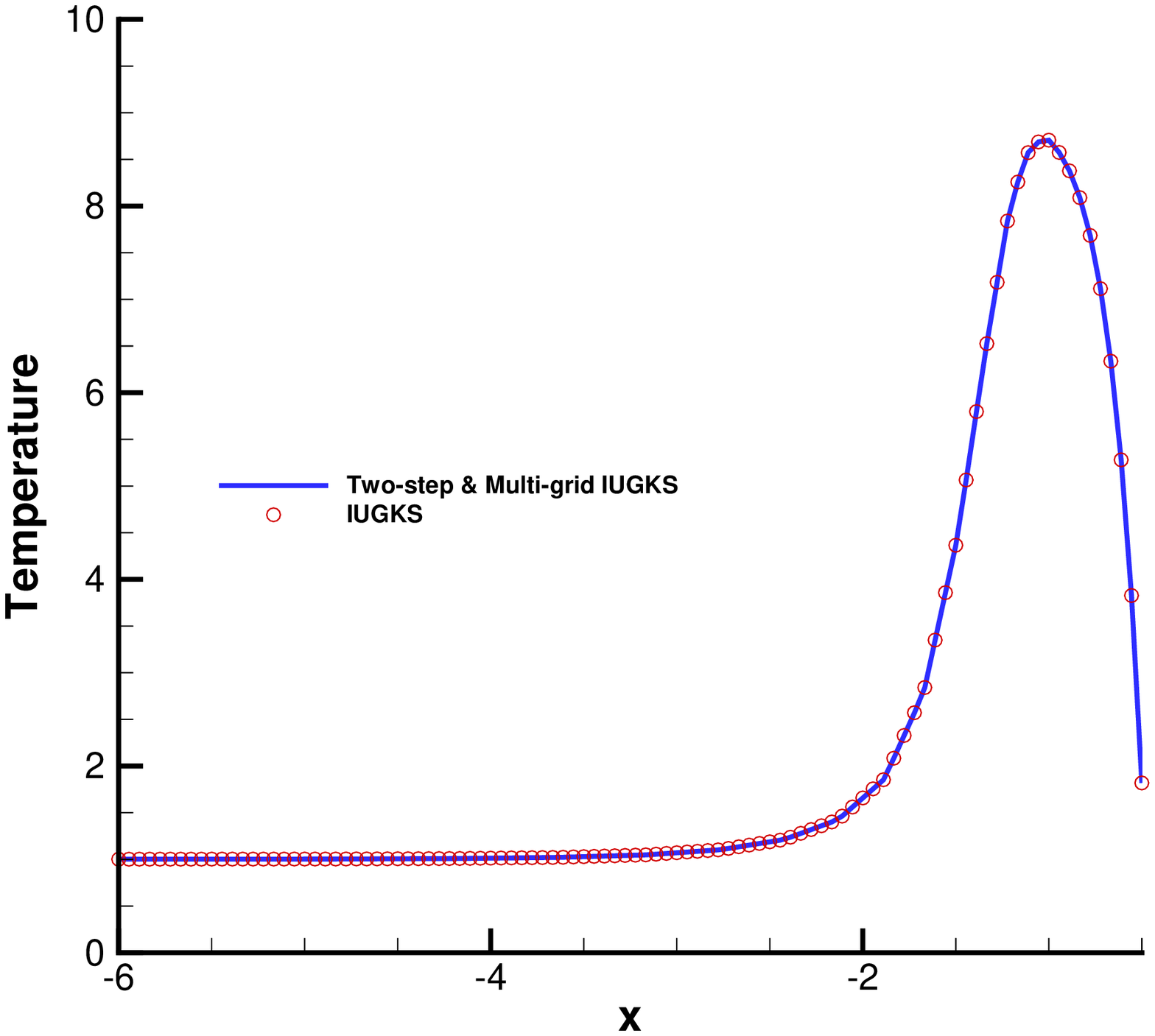}{(d)}
		\includegraphics[width=0.46\textwidth]{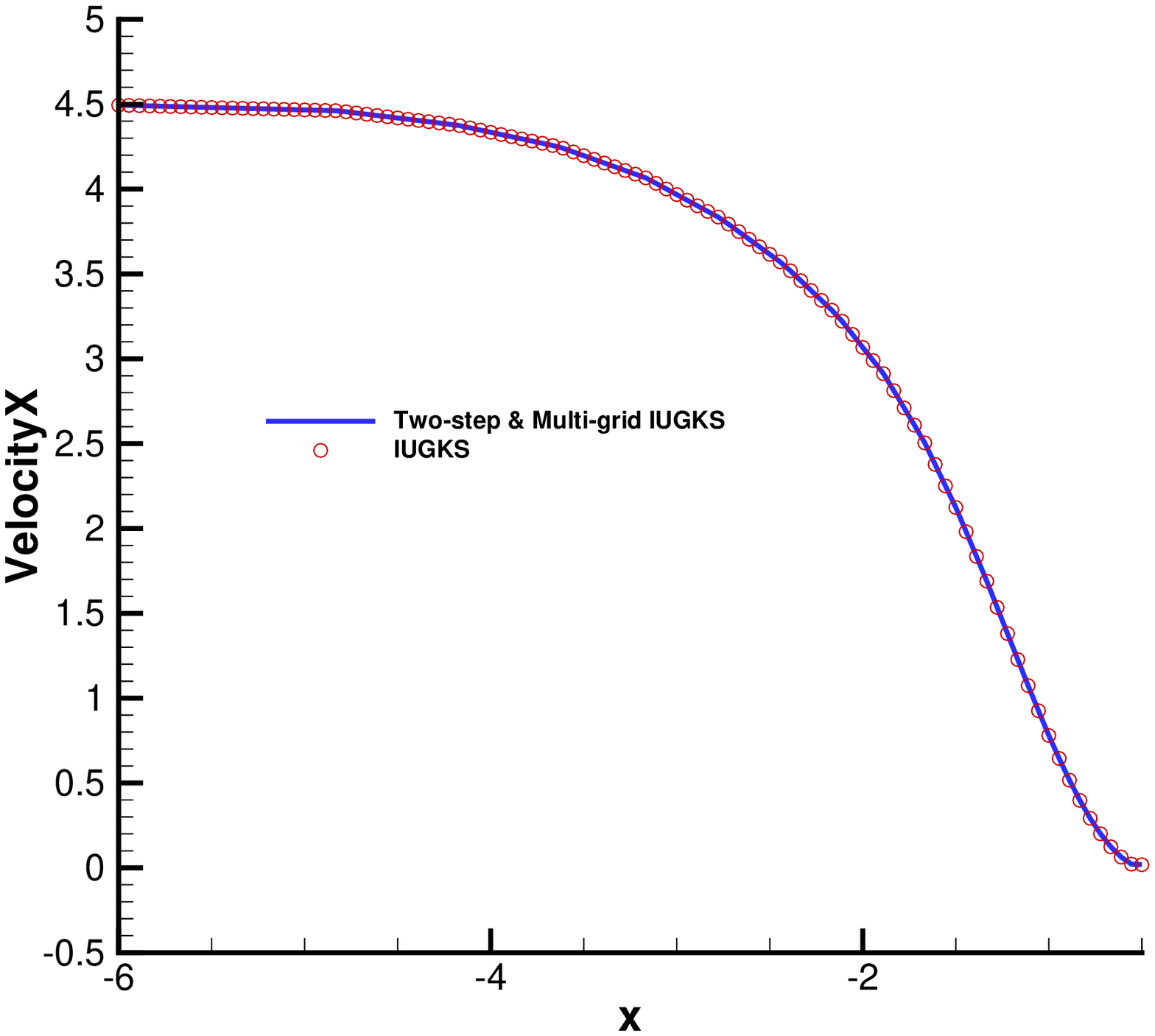}{(e)}
		\includegraphics[width=0.46\textwidth]{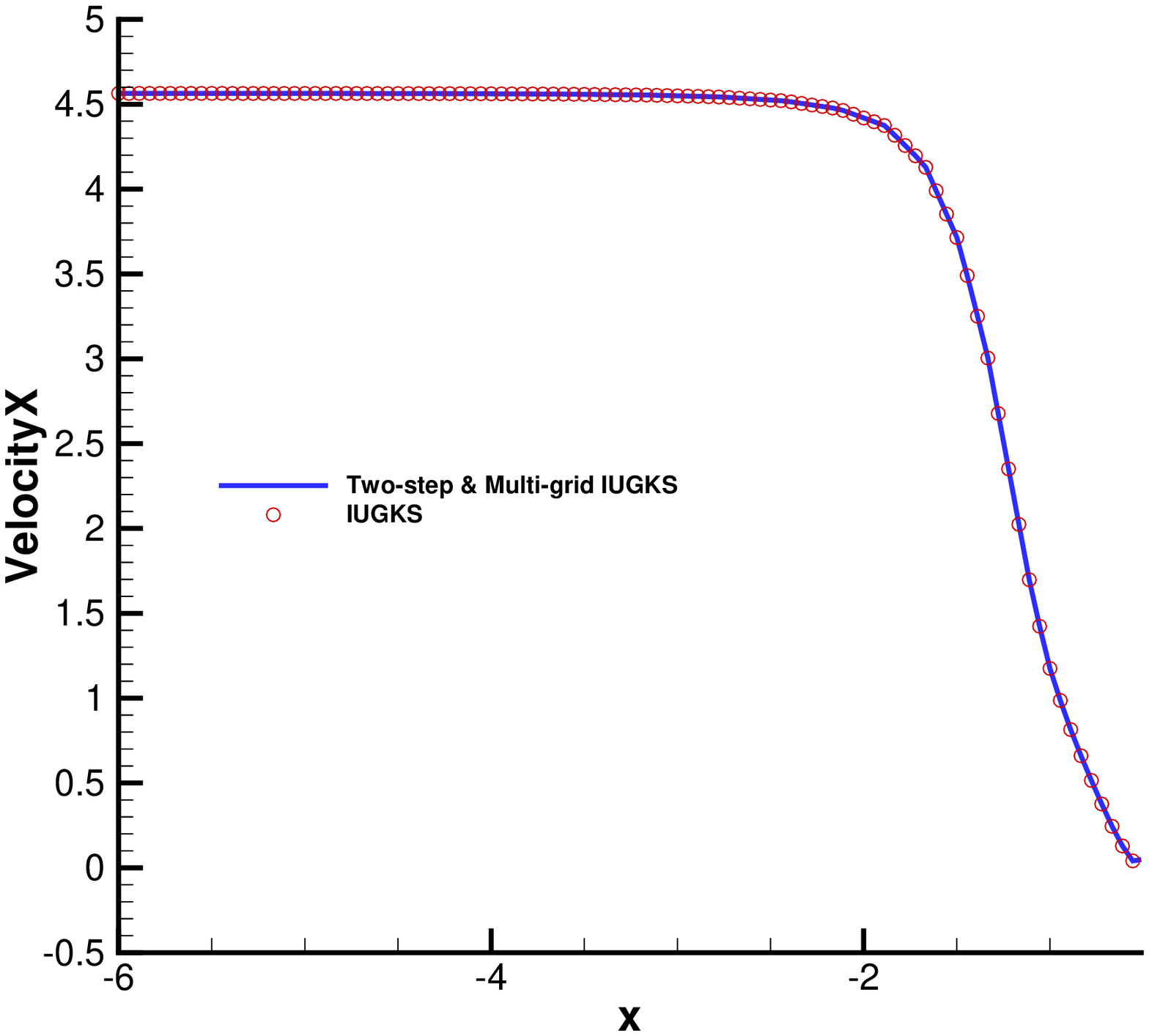}{(f)}
		\caption{Flow variables along the symmetric axis in front of the cylinder at $\mathrm{Kn} = 1$ (left) and $\mathrm{Kn} = 0.1$ (right). (a) and (b) Density; (c) and (d) temperature; (e) and (f) $U$-velocity.}
		\label{Squreline}
	\end{figure}

\begin{figure}
		\centering
		\includegraphics[width=0.46\textwidth]{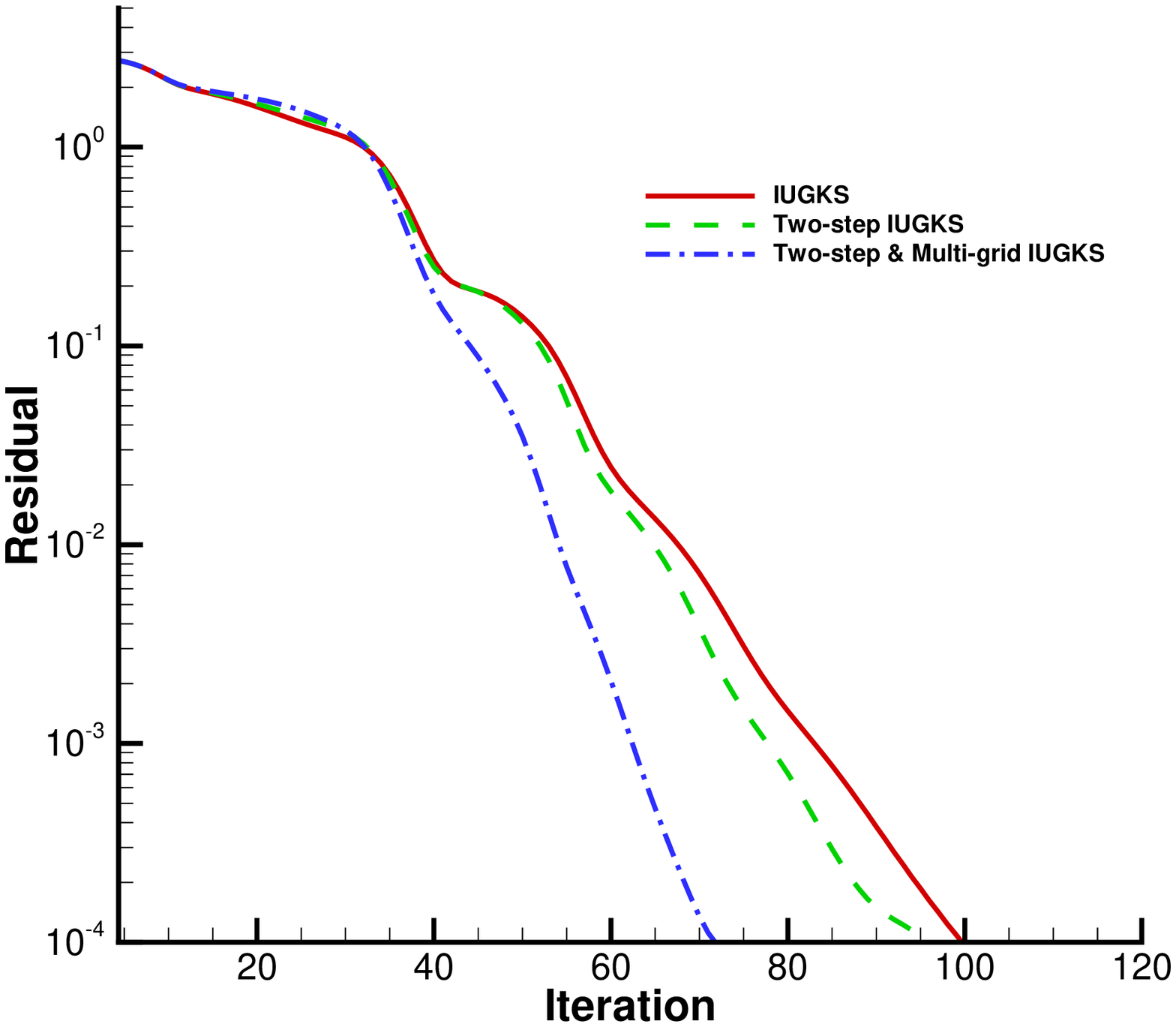}{(a)}
		\includegraphics[width=0.46\textwidth]{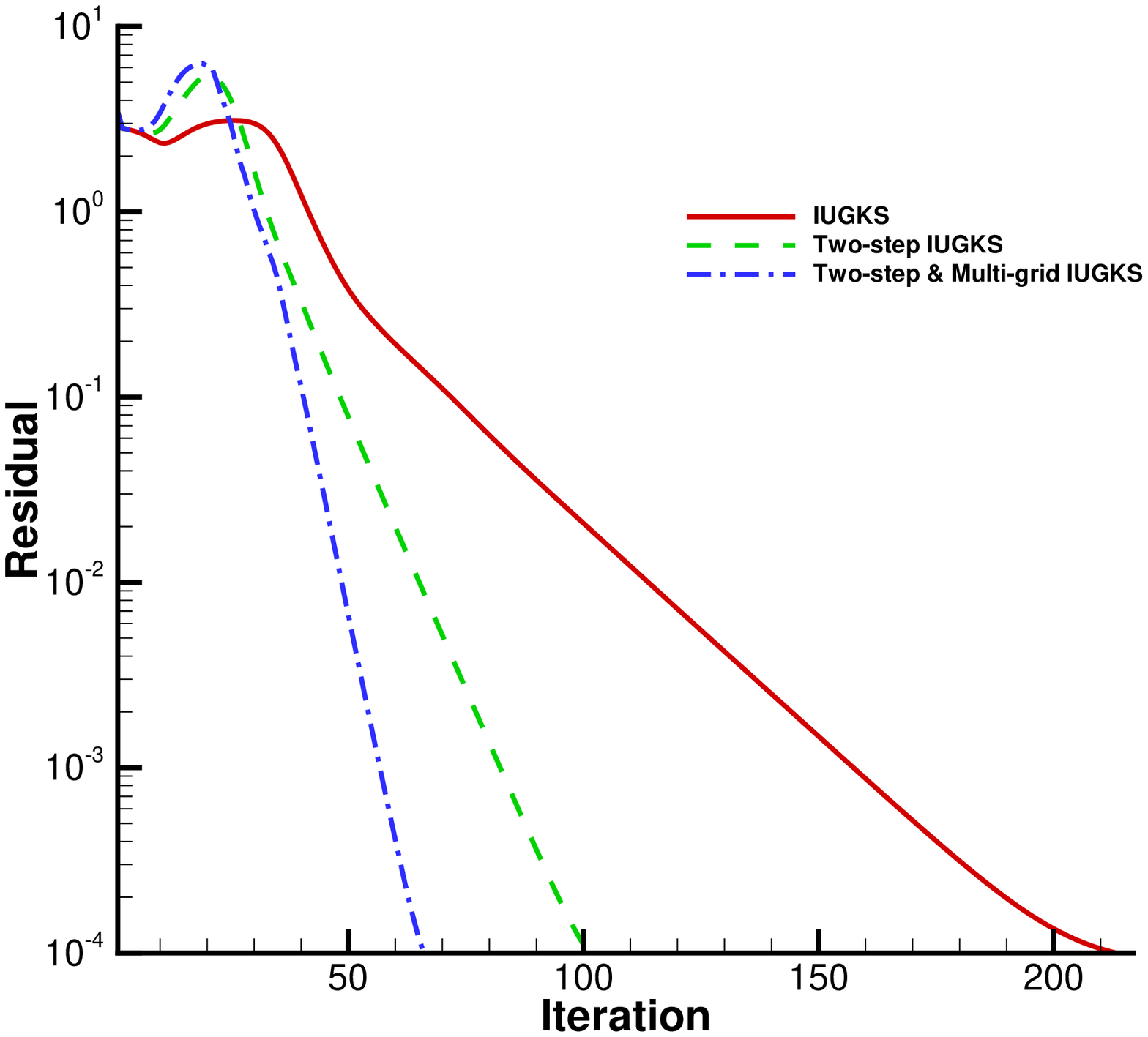}{(b)}
		\caption{Convergence history of different type of implicit UGKS in the hypersonic flow around a square cylinder. (a) $\mathrm{Kn} = 1$; (b) $\mathrm{Kn} = 0.1$;}
		\label{SqureResidual}
	\end{figure}

	The steady state solutions are presented in Fig. \ref{Squre}, and the solutions are compared with the solutions computed by the original implicit UGKS. The flow variables along the symmetric axis in the upstream are shown in Fig. \ref{Squreline}. The results from the current two-step implicit UGKS with multi-grid solver are consistent with the original IUGKS. The comparison of the convergence history and the detailed efficiency information are shown in Fig. \ref{SqureResidual} and Table. \ref{Squreeff}, respectively. Since the multi-grid method needs to do the calculations
	on both fine and coarse meshes, even with the same steps, the cost for each time step is higher than that of the other two methods.
	In the hypersonic rarefied case at $\mathrm{Kn} = 1$, the improvement in efficiency by using multi-grid method is not obvious.  In the case of $\mathrm{Kn} = 0.1$, the computational cost with multi-grid is reduced by more than two times, which is consistent with the conclusion in the cavity flow.

	\begin{table}
		\caption{Efficiency of different implicit UGKS for hypersonic flow around a square cylinder}
		\begin{tabular}{llllllllll}
			\hline
			\multirow{2}{*}{State} & \multicolumn{2}{l}{IUGKS} &  & \multicolumn{2}{l}{Two-step IUGKS} &  & \multicolumn{3}{l}{Two-step $\&$ Multi-grid IUGKS} \\ \cline{2-10}
			& Steps & Time(min) &  & Steps & Time(min) &  & Steps & Time(min) & Rate \\ \hline
			$\mathrm{Kn} = 1$   & 100    & 21.7      &  & 96    & 20.7      &  & 72    & 19.4      & 1.1 \\
			$\mathrm{Kn} = 0.1$ & 214   & 45.6      &  & 102   & 22.2      &  & 67    & 17.8      & 2.56  \\ \hline
		\end{tabular}
		\label{Squreeff}
	\end{table}
	
	\section{Conclusion}\label{conclusion}
	In this paper, a two-step implicit unified gas-kinetic scheme is constructed for the steady state solution in all flow regimes. The full Boltzmann collision operator is also integrated into the implicit scheme.
	Benefit from the coupled iterative methods for the implicit macroscopic and microscopic equations,
	the efficiency of the IUGKS can be increased significantly for steady state solution. 
By taking into account the viscous flux in the iterative matrix, the two-step acceleration technique can further improve the efficiency of the original IUGKS, especially in the near continuum and continuum flow regimes.
	In the current implicit scheme, the LU-SGS and the multi-grid methods are used to solve the algebraic systems for the macro-micro equations.
	Excellent numerical performance has been observed in all test cases, such as Couette flow, Fourier flow, cavity flow, and hypersonic flow passing over a square cylinder.
	In general, the unified gas-kinetic scheme provides a general framework. Many commonly used acceleration techniques can be implemented easily into the scheme.

	\section*{Acknowledgments}
	We would like to thank Prof. L. Wu for helpful discussion.
	The current research is supported by National Numerical Windtunnel project and  National Science Foundation of China 11772281, 91852114.

	\bibliography{xxuaybib}
	\bibliographystyle{elsarticle-num}
	\biboptions{numbers,sort&compress}
	
	\clearpage
	
\end{document}